\documentclass[useAMS, usenatbib, usegraphicx, onecolumn]{mn2e}
\usepackage{epsf}
\usepackage{graphicx}
\usepackage{dcolumn}
\usepackage{bm}
\usepackage{amssymb}
\usepackage{color}

\begin{document}

\title[The Magnetosphere of Oscillating Neutron Stars in General 
	Relativity]
      {The Magnetosphere of Oscillating Neutron Stars in General 
	Relativity}
      
\author[E.~B.~Abdikamalov, B.~J.~Ahmedov and J.~C.~Miller]
       {Ernazar~B.~Abdikamalov$^{1,\;2}$\thanks{E-mail:abdik@sissa.it},
         Bobomurat~J.~Ahmedov$^{2,\;3}$ and John~C.~Miller$^{1,\;4}$
         \\
         \\
        $^{1}$SISSA, International School for Advanced Studies, and
        INFN--Trieste, Via Beirut 2-4, 34014 Trieste, Italy     \\
        $^{2}$Institute of Nuclear Physics,
        Ulughbek, Tashkent 100214, Uzbekistan                   \\
        $^{3}$Ulugh Beg Astronomical Institute,
        Astronomicheskaya 33, Tashkent 100052, Uzbekistan       \\
        $^4$Department of Physics (Astrophysics), University of
        Oxford, Keble Road, Oxford OX1 3RH, UK
        \\}

\date{Accepted $<$date$>$. Received $<$date$>$; in original form $<$date$>$}

\pagerange{\pageref{firstpage}--\pageref{lastpage}} \pubyear{2009}

\label{firstpage}

\maketitle

\begin{abstract}
  Just as a rotating magnetised neutron star has material pulled away
  from its surface to populate a magnetosphere, a similar process can
  occur as a result of neutron-star pulsations rather than
  rotation. This is of interest in connection with the overall study
  of neutron star oscillation modes but with a particular focus on the
  situation for magnetars. Following a previous Newtonian analysis of
  the production of a force-free magnetosphere in this way
  \citet{timokhin_00}, we present here a corresponding
  general-relativistic analysis. We give a derivation of the general
  relativistic Maxwell equations for small-amplitude arbitrary
  oscillations of a non-rotating neutron star with a generic magnetic
  field and show that these can be solved analytically under the
  assumption of low current density in the magnetosphere. We apply our
  formalism to toroidal oscillations of a neutron star with a dipole
  magnetic field and find that the low current density approximation
  is valid for at least half of the oscillation modes, similarly to
  the Newtonian case. Using an improved formula for the determination
  of the last closed field line, we calculate the energy losses
  resulting from toroidal stellar oscillations for all of the modes
  for which the size of the polar cap is small. We find that general
  relativistic effects lead to shrinking of the size of the polar cap
  and an increase in the energy density of the outflowing
  plasma. These effects act in opposite directions but the net result
  is that the energy loss from the neutron star is significantly
  smaller than suggested by the Newtonian treatment. 
 \end{abstract}


\section{Introduction}
\label{introd} \noindent

Study of the internal structure of neutron stars (NSs) is of 
fundamental importance for subatomic physics since these objects 
provide a laboratory for studying the properties of high-density 
matter under very extreme conditions. In particular, there is the 
intriguing possibility of using NS oscillation modes as a probe for 
constraining models of the equation of state of matter at supranuclear 
densities. It was suggested long ago that if a NS is oscillating, 
then traces of this might be revealed in the radiation which it emits 
\citep{pacini_74, tsygan_75, boriakoff_76, bisnovatyi_95, 
ding_97, duncan_98}. Recently, a lot of interest has been focussed 
on oscillations of magnetized NSs because of the discovery of 
gamma-ray flare activity in Soft Gamma-Ray Repeaters (SGRs) which are 
thought to be the very highly magnetised NSs known as magnetars 
\citep[for recent review on the SGRs see][]{woods_06, watts_07}. 
The giant flares in these objects are thought to be powered by global 
reconfigurations of the magnetic field and it has been suggested that 
the giant flares might trigger starquakes and excite global seismic 
pulsations of the magnetar crust \citep{thompson_95, thompson_01, 
schwartz_05, duncan_98}. Indeed, analyses of the observations of 
giant flares have revealed that the decaying part of the spectrum 
exhibits a number of quasi-periodic oscillations (QPOs) with 
frequencies in the range from a few tens of Hz up to a few hundred Hz 
\citep{israel_05, strohmayer_06, watts_06} and there has been a 
considerable amount of theoretical effort attempting to identify these 
with crustal oscillation modes \citep{glampedakis_06, 
samuelsson_07, levin_07, sotani_07_a, sotani_07_b}. While 
there is substantial evidence that the observed SGR QPOs are caused by 
neutron star pulsations, there is a great deal of uncertainty about 
how stellar surface motion gets translated into the observed features 
of the X-ray radiation \citep{strohmayer_08, strohmayer_06, 
timokhin_07b}. To make progress with this, it is necessary to 
develop a better understanding of the processes occurring in the 
magnetospheres of oscillating neutron stars.

Standard pulsars typically have magnetic fields of around $10^{12}$ 
G while magnetars may have fields of up to $10^{14}-10^{15}$ G near to 
the surface. Rotation of a magnetized star generates an electric 
field:
\begin{equation}
\label{e_rot}
  E^\mathrm{rot} \sim \frac{\Omega R}{c}B \ , 
\end{equation} 
where  $ B $ is the magnetic field strength, $ c $ is the speed of
light and $ \Omega $ is the angular velocity of the star with radius $
R $. Depending on the rotation velocity and the magnetic field
strength, the electric field may be as strong as $ 10^{10} $ V
cm$^{-1}$ and it has a longitudinal component (parallel to $B$) which
can be able to pull charged particles away from the stellar surface,
if the work function is  sufficiently small, and accelerate them up to
ultra-relativistic velocities. This result led \citet{goldreich_69} to
suggest that a rotating NS with a sufficiently strong magnetic field
should be surrounded by a magnetosphere filled with charge-separated
plasma which screens the accelerating electric field and thus hinders
further outflow of charged particles from the stellar surface. Even if
the binding energy of the charged particles is sufficiently high to
prevent them being pulled out by the electric field, the NS should
nevertheless be surrounded by charged particles produced by plasma
generation processes \citep{sturrock_71,ruderman_75}, which again
screen the longitudinal component of the electric field. These
considerations led to the development of a model for pulsar
magnetospheres which is frequently called the ``standard model''
\citep[an in depth discussion and review of this can be found in,
  e.g.,][]{michel_91,beskin_93, beskin_05}.      

Timokhin, Bisnovatyi-Kogan $\&$ Spruit (2000) (referred to as TBS from 
here on) showed that an oscillating magnetized NS should also have a 
magnetosphere filled with charge-separated plasma, even if it is not 
rotating, since the vacuum electric field induced by the oscillations 
would have a large radial component which can be of the same order as 
rotationally-induced electric fields. One can show this quantitatively 
by means of the following simple arguments. To order of magnitude, the 
radial component of the vacuum electric field generated by the stellar 
oscillations is given by
\begin{equation}
  E^\mathrm{osc}\sim \frac{\omega \xi}{c}B \ ,
\end{equation}
where $\omega$ is the oscillation frequency and $\xi$ is the 
displacement amplitude. Using this together with Eq.(\ref{e_rot}), it 
follows immediately that the electric field produced by oscillations 
will be stronger than the rotationally induced one for sufficiently 
slowly-rotating neutron stars, having
\begin{equation}
  \Omega \lesssim \frac{\omega \xi }{ R } \ .
\end{equation}
 For stellar oscillations with $\xi / R \sim 0.001 $ and $\omega \sim 1$ kHz, 
the threshold is $\Omega \sim 1 $ Hz. Within this context, TBS developed a 
formalism extending the basic aspects of the standard pulsar model to the 
situation for a non-rotating magnetized NS undergoing arbitrary oscillations. 
This formalism was based on the assumption of low current densities in the 
magnetosphere, signifying that the influence of currents outside the NS on 
electromagnetic processes occurring in the magnetosphere is negligibly small 
compared to that of currents in the stellar interior. This assumption leads to 
a great simplification of the Maxwell equations, which then can be solved 
analytically. As an application of the formalism, TBS considered toroidal 
oscillations of a NS with a dipole magnetic field, and obtained analytic 
expressions for the electromagnetic field and charge density in the 
magnetosphere. (Toroidal oscillations are thought to be particularly relevant 
for magnetar QPO phenomena.) They found that the low current density 
approximation (LCDA) is valid for at least half of all toroidal oscillation 
modes and analyzed the energy losses due to plasma outflow caused by these 
modes for cases where the size of the polar cap (the region on the stellar 
surface that is crossed by open magnetic field lines) is small, finding that 
the energy losses are strongly affected by the magnetospheric plasma. For 
oscillation amplitudes larger than a certain critical value, they found that 
energy losses due to plasma outflow were larger than those due to the emission 
of the electromagnetic waves (assuming in that case that the star was 
surrounded by vacuum). Recently, \citet{timokhin_07} considered spheroidal 
oscillations of a NS with a dipole magnetic field, using the TBS formalism, 
and found that the LCDA again holds for at least half of these modes. 
Discussion in \citet{timokhin_07} also provided some useful insights into the 
role of rotation for the magnetospheric structure of oscillating NSs.

The TBS model was a very important contribution and, to the best of 
our knowledge, remains the only model for the magnetosphere of 
oscillating NSs available in the literature. However, it should be 
pointed out that it does not include several ingredients that a fully 
consistent and realistic model ought to include. Most importantly, it 
does not treat the magnetospheric currents in a fully consistent way: 
although it gives a consistent solution for around half of the 
oscillation modes, the remaining solutions turn out to be unphysical 
and, as TBS pointed out, this is a symptom of the LCDA failing there. 
Also, rotation and the effects of general relativity can be very 
relevant; in particular, several authors have stressed that using a 
Newtonian approach may not give very good results for the structure of 
NS magnetospheres \citep[see, e.g.,][]{beskin_90, muslimov_92, mofiz_00, 
morozova_08}. However, a more realistic model would naturally be more 
complicated than the TBS one whose relative simplicity can be seen as 
a positive advantage when using it as the basis for further 
applications.

The aim of the present paper is to give a general relativistic 
reworking of the TBS model so as to investigate the effects of the 
changes with respect to the Newtonian treatment. We derive the general 
relativistic Maxwell equations for arbitrary small-amplitude 
oscillations of a non-rotating spherical NS with a generic magnetic 
field configuration and show that they can be solved analytically 
within the LCDA as in Newtonian theory. We then apply this solution to 
the case of toroidal oscillations of a NS with a dipole magnetic 
field and find that the LCDA is again valid for at least half of all 
toroidal oscillation modes, as in Newtonian theory. Using an improved
formula for the determination of the last closed field line, we
calculate the energy losses resulting from these oscillations for
\textit{all} of the modes for which the size of the polar cap is small
and discuss the influence of GR effects on the energy losses. 

The paper is organized as follows. In Section 2 we introduce some definitions 
and derive the quasi-stationary Maxwell equations in Schwarzschild spacetime as 
well as the boundary conditions for the electromagnetic fields at the stellar 
surface. In Section 3 we sketch our method for analytically solving the Maxwell 
equations for arbitrary NS oscillations with a generic magnetic field 
configuration. In Section 4 we apply our formalism to the case of purely 
toroidal oscillations of a NS with a dipole magnetic field and also discuss the 
validity of the LCDA and the role of GR effects. In Section 5 we calculate the 
energy losses due to plasma outflow caused by the toroidal oscillations. Some 
detailed technical calculations related to the discussion in the main part of 
the paper are presented in Appendices A-C.

We use units for which $c = 1$, a space-like signature $(-,+,+,+)$ and 
a spherical coordinate system $(t,r,\theta ,\phi)$. Greek indices are 
taken to run from 0 to 3 while Latin indices run from 1 to 3 and we 
adopt the standard convention for summation over repeated indices. We 
indicate four-vectors with bold symbols ({\it e.g.} ${\boldsymbol u}$) 
and three-vectors with an arrow ({\it e.g.} ${\vec u}$).

\section{General Formalism}
\label{basic}

\subsection{Quasi-stationary Maxwell equations in Schwarzschild
  spacetime}

The study of electromagnetic processes related to stellar oscillations 
in the vicinity of NSs should, in principle, use the coupled system of 
Einstein-Maxwell equations. However, such an approach would be overly 
complicated for our study here, as it is for many other astrophysical 
problems. Here we simplify the problem by neglecting the contributions
of the electromagnetic fields, the NS rotation and the NS oscillations
to the spacetime metric and the structure of the NS\footnote{Several
  authors have, in fact, studied the equilibrium configurations of
  magnetars by solving the Einstein-Maxwell equations in full general
  relativity \citep{bocquet_95, bonazzola_96, cardall_01} or by using
  perturbative techniques \citep{colaiuda_08, haskell_08}.}, noting
that this is expected to be a good approximation for small-amplitude
oscillations. Indeed, for a star with average mass-energy density $
\bar \rho $, mass $M$ and radius $R$, the maximum fractional change in
the spacetime metric produced by the magnetic field is typically of
the same order as the ratio between the energy density in the surface
magnetic field and average mass-energy density of the NS, i.e., 
\begin{equation}
\label{eqq:metric_corrections}
\frac{B^2} {8 \pi \bar \rho c^2} \simeq 10^{-7} \left(
\frac{B} {10^{15} ~ \mathrm{G}} \right)^2 \left(\frac{1.4 ~ M_{\odot}}
     {M} \right) \left( \frac{R} {10 ~ \mathrm{km}} \right)^3 \ .
\end{equation}
The corresponding fractional change in the metric due to rotation is
of order
\begin{equation}
0.1 \left( \frac{\Omega }{ \Omega_\mathrm{K}} \right)^2 = 10^{-7}
\left( \frac{ \Omega }{ 1 ~ \mathrm{Hz}} \right)^2 \left(
\frac{ 1 ~ \mathrm{kHz} } {\Omega_\mathrm{K} } \right)^2 
\end{equation}
where $ \Omega_\mathrm{K} $ is the Keplerian angular velocity at the
surface of the NS. Moreover, in the case of magnetars, which we
consider in our study, the oscillations are thought to be triggered by
the global reconfiguration of the magnetic field. Due to this reason, the
corrections due to the oscillations should not exceed the contribution
due to the magnetic field itself given by estimate
(\ref{eqq:metric_corrections}). Therefore, we can safely work in the
background spacetime of a static spherical star, whose line element in
a  spherical coordinate system $(t,r,\theta ,\phi)$ is given by
\begin{equation} \label{schw_g} 
ds^2 = g_{00}(r) dt^2 + g_{11}(r) dr^2+
        r^2 d\theta ^2+ r^2\sin^2\theta d\phi ^2 \ ,
\end{equation}
 while the geometry of the spacetime external to the star (i.e. for $r 
\ge R$) is given by the Schwarzschild solution:
\begin{equation}
\label{schw} ds^2 = -N^2 dt^2 + N^{-2} dr^2+r^2 d\theta ^2
 + r^2\sin^2\theta d\phi ^2 \ ,
\end{equation}
 where $N \equiv (1 - 2M/r)^{1/2}$ and $M$ is the total mass of the 
star. For the part of the spacetime inside the star, we represent the 
metric in terms of functions $\Lambda$ and $\Phi$ as
\begin{equation}
g_{00} = -e^{2 \Phi(r)} \ , \hskip 0.5 cm g_{11} = e^{2
\Lambda(r)} = \left(1-\frac{2 m(r)}{r}\right)^{-1}\ ,
\end{equation}
 where $m(r)=4\pi\int^r_0 r'^2 \rho(r') dr'$ is the volume integral of the 
total energy density $\rho(r)$ over the spatial coordinates. The form of these 
functions is given by solution of the standard TOV equations for spherical 
relativistic stars \citep[see, e.g.,][]{shapiro_83} and they are matched 
continuously to the external Schwarzschild spacetime through the relations
\begin{equation}
g_{00}(r=R) = N_{_R}^2 \ , \hskip 0.5 cm g_{11}(r=R) = N_{_R}^{-2} \ ,
\end{equation}
 where $N_{_{\rm R}} \equiv (1 - 2M/R)^{1/2}$. Within the external
part of the spacetime, we select a family of static observers with
four-velocity components given by
\begin{equation}
\label{obs}
(u^{\alpha})_{_{\rm obs}}\equiv
        N^{-1}\bigg(1,0,0,0\bigg) \ .
\end{equation}
 and associated orthonormal frames having tetrad four
vectors $\{{\textbf{e}}_{\hat \mu}\} = ({\boldsymbol e}_{\hat 0},
{\boldsymbol e}_{\hat r}, {\boldsymbol e}_{\hat \theta},
{\boldsymbol e}_{\hat \phi})$ and 1-forms $\{{\boldsymbol
\omega}^{\hat \mu}\} = ({\boldsymbol \omega}^{\hat 0},
{\boldsymbol \omega}^{\hat r}, {\boldsymbol \omega}^{\hat \theta},
{\boldsymbol \omega}^{\hat \phi})$,  which will become useful when
determining the ``physical'' components of the electromagnetic
fields. The components of the vectors are given by equations (6)-(9)
of \citet{rezzolla_04} (hereafter Paper I).

The general relativistic Maxwell equations have the following form
\citep{landau_87}
\begin{equation}
\label{maxwell_firstpair}
3 F_{[\alpha \beta, \gamma]} = F_{\alpha \beta, \gamma }
        + F_{\gamma \alpha, \beta} + F_{\beta \gamma,\alpha}
        = 0 \ ,
\end{equation}
\begin{equation}
\label{maxwell_secondpair}
F^{\alpha \beta}_{\ \ \ \ ;\beta} = 4 \pi J^{\alpha} \ ,
\end{equation}
where $ F^{\alpha \beta} $ is the electromagnetic field tensor and
${\boldsymbol J}$ is the electric-charge 4-current. We consider the
region close to the star (the near zone), at distances from the NS
much smaller than the wavelength $ \lambda = 2 \pi c / \omega $. In
the near zone the electromagnetic fields are quasi-stationary,
therefore we neglect the displacement current term in the Maxwell
equations. Once expressed in terms of the physical components of the
electric and magnetic fields, equations (\ref{maxwell_firstpair}) and
(\ref{maxwell_secondpair}) become (see Section 2 of Paper I for
details of the derivation) 
\begin{eqnarray}
\label{max1a} \sin\theta \partial_r\left(r^2B^{\hat r}\right)+
        N^{-1}r\partial_\theta\left(\sin\theta B^{\hat \theta}\right) +
        N^{-1} r \partial_\phi B^{\hat \phi} = 0 \ ,
\end{eqnarray}
\begin{eqnarray}
\label{max1b} \left({r\sin\theta}\right)\frac{\partial B^{\hat
r}}{\partial t}  &=&  N \left[\partial_\phi E^{\hat\theta}-
       \partial_\theta\left(\sin\theta
        E^{\hat \phi} \right)\right] \ ,
\\
\label{max1c} \left({N^{-1}r\sin\theta}\right)
        \frac{\partial B^{\hat \theta}}{\partial t}
        &=& - \partial_\phi E^{\hat r} +
        \sin\theta \partial_r\left(r N E^{\hat \phi} \right)
        \ ,
\\
\label{max1d} \left({N^{-1}r}\right)
        \frac{\partial B^{\hat \phi}}{\partial t}
        &=& - \partial_r\left(r N E^{\hat \theta}\right)
        + \partial_\theta E^{\hat r} \ ,
\end{eqnarray}
\begin{eqnarray}
\label{max2a}
N \sin\theta\partial_r\left(r^2 E^{\hat r} \right)+
        {r}\partial_\theta\left(\sin\theta E^{\hat \theta}\right)
        + r \partial_\phi E^{\hat \phi}
        = {4\pi\rho_e}r^2\sin\theta \ , &&
\end{eqnarray}
\begin{eqnarray}
\label{max2b}
\left[\partial_\theta\left(\sin\theta  B^{\hat \phi} \right)
        - \partial_\phi B^{\hat\theta}\right]
        & = & {4\pi}r\sin\theta J^{\hat r} \ ,
\\
\label{max2c}
\partial_\phi B^{\hat r} - \sin\theta \partial_r\left(rN
        B^{\hat \phi} \right)
        & = & {4\pi}r\sin\theta J^{\hat\theta} \ ,
\\
\label{max2d}
\partial_r\left(Nr B^{\hat \theta} \right) -
        \partial_\theta B^{\hat r}
        & = &
        {4\pi}rJ^{\hat\phi} \ ,
\end{eqnarray}
where $\rho_e$ is the proper charge density. We further assume that
the force-free condition,
\begin{equation}
\label{force_free} \vec{E}_{_{\rm SC}}\cdot \vec{B}=0 \ ,
\end{equation}
is fulfilled everywhere in the magnetosphere, implying that the
magnetosphere of the NS is populated with charged particles that
cancel the longitudinal component of the electric field. The charge
density $ \rho_{_{\rm SC}} $ responsible for the electric field $
\vec{E}_{_{\rm SC}} $ (cf. equation \ref{max2a}) is the characteristic
charge density of the force-free magnetosphere; this is appropriate
for describing the charge density in  the inner parts of the NS
magnetosphere. We will refer to $ \vec{E}_{_{\rm SC}} $ as the
space-charge (SC) electric field, while to $ \rho_{_{\rm SC}} $ as the
SC charge density.

Finally, we introduce the perturbation of the NS crust in terms of its
four-velocity, with the components being given by
\begin{equation}
\label{vel} w^\alpha \equiv e^{-\Phi}
        \bigg(1,\frac{dx^i}{dt}\bigg)=
        e^{-\Phi}\bigg(1,e^{-\Lambda}\delta v^{\hat r},\frac{\delta v^{\hat\theta}}{r},
        \frac{\delta v^{\hat\phi}}{r\sin\theta}\bigg) \ ,
\end{equation}
where $\delta v^i=dx^i/dt$ is the relative oscillation three-velocity
of the conducting stellar surface with respect to the unperturbed
state of the star.

\subsection{Boundary conditions at the surface of star}

We now begin our study of the internal electromagnetic field induced 
by the stellar oscillations. We assume here that the material in the 
crust can be treated as a perfect conductor and the induced electric 
field then depends on the magnetic field and the pulsational velocity 
field according to the following relations (see Paper I for details of 
the derivation):
\begin{eqnarray}
\label{in_ef2}
&&E_{in}^{\hat r} =
        -e^{-\Phi}\left[\delta v^{\hat\theta} B^{\hat\phi}
        -\delta v^{\hat\phi}B^{\hat \theta}\right] \ ,
\\\nonumber\\
\label{in_ef3} &&E^{\hat\theta}_{in} = - e^{-\Phi}\left[\delta
v^{\hat\phi} B^{\hat r}
        -\delta v^{\hat r}B^{\hat \phi}\right] \ ,
\\\nonumber\\
\label{in_ef4} &&E_{in}^{\hat\phi} = - e^{-\Phi}\left[\delta
v^{\hat r} B^{\hat\theta}
        -\delta v^{\hat \theta}B^{\hat r}\right] \ .
\end{eqnarray}

Boundary conditions for the magnetic field at the stellar surface 
$(r=R)$ can be obtained from the requirement of continuity for the 
radial component, while leaving the tangential components free to be 
discontinuous because of surface currents:
\begin{eqnarray}
\label{bound_b1}
&&B^{\hat r}_{ex}|_{r=R}=B^{\hat r}_{in}|_{r=R}\ ,
\\\nonumber\\
\label{bound_b2}
&&B^{\hat\theta}_{ex}|_{r=R}=B^{\hat\theta}_{in}|_{r=R}
    +4\pi i^{\hat\phi}\ ,
\\\nonumber\\
\label{bound_b3}
&&B^{\hat\phi}_{ex}|_{r=R}=B^{\hat\phi}_{in}|_{r=R}
    -4\pi i^{\hat\theta}\ ,
\end{eqnarray}
 where $i^{\hat i}$ is the surface current density. Boundary 
conditions for the electric field at the stellar surface are obtained 
from requirement of continuity of the tangential components, leaving 
$E^{\hat r}$ to have a discontinuity proportional to the surface 
charge density $\Sigma_{s}$:
\begin{eqnarray}
\label{bound_e1} E^{\hat r}_{ex}|_{r=R} & =& E^{\hat r}_{in}|_{r=R}
+4\pi\Sigma_s  =  -N^{-1}_R \left[\delta v^{\hat\theta}
  B^{\hat\phi}  -\delta v^{\hat\phi}B^{\hat
    \theta}\right]|_{r=R}+4\pi\Sigma_s \ ,
\\\nonumber\\
\label{bound_e2}
E^{\hat\theta}_{ex}|_{r=R} & = & E^{\hat\theta}_{in}|_{r=R}= -
N^{-1}_R\left[\delta v^{\hat\phi} B^{\hat r}
        -\delta v^{\hat r}B^{\hat \phi}\right]|_{r=R} \ ,
\\\nonumber\\
\label{bound_e3}
E^{\hat\phi}_{ex}|_{r=R} & = & E^{\hat\phi}_{in}|_{r=R}= -
N^{-1}_R\left[\delta v^{\hat r} B^{\hat\theta}
        -\delta v^{\hat \theta}B^{\hat r}\right]|_{r=R} \ ,
\end{eqnarray}
where $\Sigma_s$ is the surface charge density.

\subsection{The low current density approximation}
\label{lcda}

The low current density approximation was introduced by TBS, and in 
the present section we present a brief introduction to it for 
completeness. Close to the NS surface, the current flows along the 
magnetic field lines, and so in the inner parts of the magnetosphere 
it can be expressed as
\begin{equation}
\label{current}
\vec{J} = \alpha(r,\theta,\phi) \cdot \vec{B}\ ,
\end{equation}
 where $\alpha$ is a scalar function. The system of equations 
(\ref{max1a})--(\ref{max2d}), (\ref{force_free}) and (\ref{current}) 
forms a complete set but is overly complicated for solving in the 
general case. However, within the LCDA these equations can, as we show 
below, be solved analytically for arbitrary oscillations of a NS with 
a generic magnetic field configuration.

The LCDA scheme is based on the assumption that the perturbation of 
the magnetic field induced by currents flowing in the NS interior is 
much larger than that due to currents in the magnetosphere, which are 
neglected to first order in the oscillation parameter 
$\bar{\xi}\equiv\xi/R$:
\begin{equation}
\frac{4\pi}{c}\vec J \ll \nabla \times \vec B \ ,
\end{equation}
and 
\begin{equation}
 \nabla \times \vec{B}^{(1)} = 0 \ , 
\end{equation}
 where $\vec B^{(1)}$ is the first order term of the expansion in $ 
\bar{\xi} $. This also implies that the current density satisfies the 
condition
\begin{equation}
\label{lca}
J\ll \frac{1}{r}\left(B_{(0)}\ \frac{\xi}{R}\right) c \approx
\rho_{_{\rm SC}}(R) \ c \left(\frac{c}{\omega r}\right)\ ,
\end{equation}
 where $\rho_{_{\rm SC}}(R)$ is the SC density near to the surface of 
the star. Here we have used the relation $\rho_{_{\rm SC}} (R) \simeq 
B^{(0)} \eta / c R $, where $\eta$ is the velocity amplitude of the 
oscillation and $\omega$ is its frequency.

In regions of complete charge separation, the maximum current density 
is given by $\rho_{_{\rm SC}}c$. Since the absolute value of 
$\rho_{_{\rm SC}}$ decreases with increasing $r$ and because $ r \ll c
/ w $ in the near zone, condition (\ref{lca}) is satisfied in the
magnetosphere if there is complete charge separation there. Since the
current in the magnetosphere flows along magnetic field lines, its
magnitude does not change and so condition (\ref{lca}) is also
satisfied along magnetic field lines in non-charge-separated regions
as long as they have crossed regions with complete charge separation. 

In the following, we solve the Maxwell equations assuming that 
condition (\ref{lca}) is satisfied throughout the whole near zone. As 
discussed above, a regular solution of the system of equations 
(\ref{max1a})-(\ref{max2d}), (\ref{force_free}) and (\ref{current}) 
should exist for arbitrary oscillations and arbitrary configurations 
of the NS magnetic field and so, as shown by TBS, if a solution has an 
unphysical behaviour, this would imply that the LCDA fails for this 
oscillation and that the accelerating electric field cannot be 
screened only by a stationary configuration of the charged-separated 
plasma. In some regions of the magnetosphere, the current density 
could be as high as
\begin{equation}
\label{cur_density} J\simeq\rho_{_{\rm SC}}c\left(\frac{c}{\omega
r}\right)\ .
\end{equation}
 For a more detailed discussion of the LCDA and its validity, we refer 
the reader to Sections 2.3 and 3.2.1 of TBS.

\section{The LCDA solution}

\subsection{The electromagnetic field in the magnetosphere}

We now begin our solution of the Maxwell equations, assuming that the 
LCDA condition (\ref{lca}) is satisfied everywhere in the 
magnetosphere. Within the LCDA, equations (\ref{max2b})-(\ref{max2d}) 
for the magnetic field in the magnetosphere take the form
\begin{eqnarray}
\label{max2b_m}
\partial_\theta\left(\sin\theta  B^{\hat \phi} \right)
        - \partial_\phi B^{\hat\theta}
        & = & 0 \ ,
\\
\label{max2c_m}
\partial_\phi B^{\hat r} - \sin\theta \partial_r\left(rN
        B^{\hat \phi} \right)
        & = & 0 \ ,
\\
\label{max2d_m}
\partial_r\left(Nr B^{\hat \theta} \right) -
        \partial_\theta B^{\hat r}
        & = & 0 \ .
\end{eqnarray}
 As demonstrated in Paper I, the components of the magnetic field
$B^{\hat r}, B^{\hat \theta}$ and $B^{\hat \phi}$ can be expressed in
terms of a scalar function $S$ in the following way:
\begin{eqnarray}
\label{ex_mf1}  B^{\hat r} &=&
-\frac{1}{r^2\sin^2\theta}\left[\sin\theta\partial_\theta
      \left(\sin\theta\partial_\theta S\right)+
      \partial_\phi\partial_\phi S\right]\ ,
\\\nonumber\\
\label{ex_mf2}
B^{\hat\theta} &=& \frac{N}{r}\partial_\theta\partial_r S\ ,
\\\nonumber\\
\label{ex_mf3}
B^{\hat\phi} &=& \frac{N}{r\sin\theta}\partial_\phi\partial_r S \ .
\end{eqnarray}
 Substituting these expressions into the Maxwell equations 
(\ref{max1b})--(\ref{max1d}), we obtain a system of equations for the 
electric field components which has the following general solution
\begin{eqnarray}
\label{ex_ef1} && E^{\hat r}_{_{\rm SC}}=-\partial_r(\Psi_{_{\rm
SC}}) \ ,
\\\nonumber\\
\label{ex_ef2} && E^{\hat \theta}_{_{\rm
SC}}=-\frac{1}{Nr\sin\theta}\partial_t\partial_\phi S
    -\frac{1}{Nr}\partial_\theta(\Psi_{_{\rm SC}})\ ,
\\\nonumber\\
\label{ex_ef3} && E^{\hat \phi}_{_{\rm
SC}}=\frac{1}{Nr}\partial_t\partial_\theta S
    -\frac{1}{Nr\sin\theta}\partial_\phi
    \left(\Psi_{_{\rm SC}}\right) \ ,
\end{eqnarray}
 where $\Psi_{_{\rm SC}}$ is an arbitrary scalar function. The terms 
proportional to the gradient of $\Psi_{_{\rm SC}}$ are responsible for 
the contribution of the charged particles in the magnetosphere. The 
vacuum part of the electric field is given by the derivatives of the 
scalar function $S$. Substituting (\ref{ex_ef1})-(\ref{ex_ef3}) into 
equation (\ref{max2a}), we get an expression for the SC charge density 
in terms of $\Psi_{_{\rm SC}}$:
\begin{equation}
\label{SC_charge} \rho_{_{\rm SC}}= -\frac{1}{4\pi
      r^2}\left[N\partial_r\left(r^2\partial_r\Psi_{SC}
      \right)+\frac{1}{N}\triangle_{\Omega}\Psi_{SC}\right]\,,
\end{equation}
 where $\triangle_{\Omega}$ is the angular part of the Laplacian:
\begin{equation}
 \triangle_{\Omega}=\frac{1}{\sin\theta}\partial_{\theta}
 \left(\sin\theta\partial_{\theta}\right)+
 \frac{1}{\sin^2\theta}\partial_{\phi\phi}.
\end{equation}
%

\subsection{The equation for $\Psi_{_{\rm SC}}$}

Substituting expressions (\ref{ex_mf1})--(\ref{ex_mf3}) and 
(\ref{ex_ef1})--(\ref{ex_ef3}) for the components of the electric and 
magnetic fields into the force-free condition (\ref{force_free}), we 
get the following equation for $\Psi_{_{\rm SC}}$
\begin{eqnarray}
\label{Psi}
\frac{1}{\sin^2\theta}\left[\sin\theta\partial_\theta
  \left(\sin\theta\partial_\theta
    S\right)+\partial_\phi\partial_\phi S\right]
    \partial_r(\Psi_{_{\rm SC}}) & - & \frac{1}{\sin\theta} \left[
      \partial_\phi \partial_tS \partial_\theta\partial_rS  -
    \partial_\theta\partial_tS \partial_\phi\partial_rS\right]
    \nonumber \\ \nonumber \\ & - & \partial_\theta\partial_rS
    \partial_\theta (\Psi_{_{\rm SC}}) - \frac{1}{\sin^2\theta}
    \partial_\phi \partial_rS \partial_\phi \left( \Psi_{_{\rm SC}}
    \right) = 0 \ .
\end{eqnarray}
 If the amplitude of the NS oscillations is suitably small ($\bar{\xi} 
\ll 1$), the function $S$ can be series expanded in terms of the 
dimensionless perturbation parameter $ \bar{\xi} $ and can be 
approximated by the sum of the two lowest order terms
\begin{equation}
\label{exp_P2} S(t,r,\theta,\phi) = S_0(r,\theta,\phi)+
    \delta S(t,r,\theta,\phi)\ .
\end{equation}
 Here the first term $S_0$ corresponds to the unperturbed static 
magnetic field of the NS, while $\delta S$ is the first order 
correction to it. At this level of approximation, equation (\ref{Psi}) 
for $\Psi_{_{\rm SC}}$ takes the form
\begin{eqnarray}
\label{Psi_0}
\frac{1}{\sin^2\theta} \left[\sin\theta\partial_\theta
    \left(\sin\theta \partial_\theta S_{0}\right) + \partial_\phi
    \partial_\phi S_{0}\right] \partial_r(\Psi_{_{\rm SC}})
    & - & \frac{1}{\sin\theta} \left[\partial_\phi\partial_t(\delta S)
      \partial_\theta\partial_r S_{0} -
    \partial_\theta\partial_t(\delta S)\partial_\phi\partial_r
    S_{0}\right]
\nonumber\\ \nonumber\\
 &- &\partial_\theta\partial_r S_{0}
    \partial_\theta (\Psi_{_{\rm SC}})-
    \frac{1}{\sin^2\theta}\partial_\phi\partial_r S_{0}
    \partial_\phi\left(\Psi_{_{\rm SC}}\right)=0\ .
\end{eqnarray}
 Next we expand $S$ in terms of the spherical harmonics:
\begin{equation}
\label{exp_P} S=\sum_{\ell=0}^{\infty}\sum_{m=-\ell}^{\ell}
S_{\,\ell m}(t,r)Y_{\ell m}(\theta,\phi )\ .
\end{equation}
 where the functions $S_{\,\ell m}$ are given in terms of Legendre 
functions of the second kind $Q_{\ell}$ by \citep{rezzolla_01}
\begin{equation}
\label{sol_P} {S}_{\ell m}(t,r)=-
\frac{r^2}{M^2}\frac{d}{dr}\left[r\left(1-\frac{2M}{r}\right)
    \frac{d}{dr}Q_{\ell}\left(1-\frac{r}{M}\right)\right]
s_{\ell m}(t)\ .
\end{equation}
 Note that all of the time dependence in (\ref{sol_P}) is contained in 
the integration constants $s_{\ell m}(t)$ which, as we will see later, 
are determined by the boundary conditions at the surface of the star. 
We now series expand the coefficients $S_{\ell m}(t,r)$ and $s_{\ell 
m}(t)$ in terms of $\bar{\xi}$
\begin{equation}
S_{\ell m}(t,r) = S_{0\ell m}(r)+\delta S_{\ell m}(t,r),
\qquad
s_{\ell m}(t)  = s_{0\ell m}+\delta s_{\ell m}(t) \ ,
\end{equation}
 where all of the time dependence is now confined within the 
coefficients $\delta S_{\ell m}(r,t)$ and $\delta s_{\ell m}(t)$, 
while the coefficients $S_{0\ell m}$ and $s_{0\ell m}$ are responsible 
for the unperturbed static magnetic field of the star. Using these 
results, we can also express $S$ and $\delta S$ in terms of a series 
in $Y_{\ell m} (\theta, \phi) $ in the following way
\begin{eqnarray}
  \label{exp_S0}
  S_0&=&\sum_{\ell=0}^{\infty}\sum_{m=-\ell}^{\ell} S_{0\,\ell m} (r)
  Y_{\ell m} (\theta,\phi ), \ \\\nonumber\\
  \label{exp_dS}
  \delta S & =& \sum_{\ell=0}^{\infty} \sum_{ m=-\ell }^{\ell} \delta
    S_{\,\ell m}(t,r)Y_{\ell m}(\theta,\phi )\ .
\end{eqnarray}
 The variables $r$ and $t$ in the functions ${S}_{\ell m}(t,r)$ and 
${S}_{\ell m}(t,r)$ can be separated using relation (\ref{sol_P}):
\begin{eqnarray}
\label{sol_P1}
      {S}_{0\ell m}(r) &=& - \frac{r^2}{M^2} \frac{d}{dr}
    \left[r\left(1-\frac{2M}{r}\right) \frac{d}{dr} Q_{\ell}
    \left(1-\frac{r}{M}\right)\right]  s_{0\ell  m} \ ,
    \\\nonumber\\
\label{sol_P2}
    {S}_{\ell m}(t,r) &=& - \frac{r^2}{M^2} \frac{d}{dr} \left[ r
    \left( 1 - \frac{ 2M }{r} \right) \frac{d}{dr} Q_{\ell} \left(1 -
    \frac{r}{M}\right)\right]  \delta s_{\ell  m}(t)\ .
\end{eqnarray}
%

\subsection{The boundary condition for $ \Psi_{_{\rm SC}} $}

We now derive a boundary condition for $\Psi_{_{\rm SC}}$ at the 
stellar surface using the behaviour of the electric and magnetic 
fields in that region. Following TBS, we assume that near to the
stellar surface the interior magnetic field has the same behaviour as
the exterior one:
\begin{eqnarray}
\label{in_mf1}
&&B^{\hat r}=-\frac{C_1}{r^2\sin^2\theta}
    \left[\sin\theta\partial_\theta\left(\sin\theta
    \partial_\theta S\right)+
    \partial_{\phi\phi} S\right]\ ,
\\\nonumber\\
\label{in_mf2}
&&B^{\hat\theta}=C_1\frac{e^{-\Lambda}}{r}\partial_\theta
     \partial_r S\ ,
\\\nonumber\\
\label{in_mf3}
&&B^{\hat\phi}=C_1\frac{e^{-\Lambda}}{r\sin\theta}
      \partial_\phi\partial_r S\ .
\end{eqnarray}

Using the continuity condition for the normal component of the 
magnetic field $\left[B^{\hat r}\right]=0$ at the stellar surface
\citep{pons_07} together with the condition $e^{-\Lambda}|_{r=R}\equiv
N_R$, one finds that the integration constant $C_1$ is equal to
one. The interior electric field components can then be obtained by
substituting (\ref{in_mf1}) -- (\ref{in_mf3}) (with $C_1 = 1$) into
(\ref{in_ef2}) -- (\ref{in_ef4}):
\begin{eqnarray}
\label{int_ef1}
E^{\hat r}_{in} &=& -\frac{e^{-(\Phi+\Lambda)}}{r\sin\theta}
    \left\{\delta v^{\hat\theta}\partial_\phi\partial_rS-
    \sin\theta \delta v^{\hat\phi}\partial_\theta\partial_rS\right\}\ ,
\\\nonumber\\
\label{int_ef2}
E^{\hat \theta}_{in} &=& \frac{e^{-(\Phi+\Lambda)}}{r\sin\theta}
    \left\{\delta v^{\hat r}\partial_\phi\partial_rS+
    \frac{\delta v^{\hat\phi}e^{\Lambda}}{r\sin\theta}
    \left[\sin\theta\partial_\theta\left(\sin\theta
    \partial_\theta S\right)+\partial_\phi\partial_\phi S\right]
    \right\}\ ,
\\\nonumber\\
\label{int_ef3}
E^{\hat \phi}_{in} &=&
    -\frac{e^{-(\Phi+\Lambda)}}{r}
    \left\{\delta v^{\hat r}\partial_\theta\partial_r S+
    \frac{\delta v^{\hat\theta}e^{\Lambda}}{r\sin^2\theta}
    \left[\sin\theta\partial_\theta\left(\sin\theta
    \partial_\theta S\right)+\partial_\phi\partial_\phi S
    \right] \right\}\ .
\end{eqnarray}

The continuity condition for the $\theta$ component of the electric 
field across the stellar surface (\ref{bound_e2}) gives a boundary 
condition for $\partial_\theta \Psi_{_{{\rm SC}}}|_{r=R}$:
\begin{equation}
\label{bound_Psi_2}
\Psi_{_{{\rm SC},\theta}}|_{r=R} =
    -\left\{\frac{\delta v^{\hat\phi}}{R\sin^2\theta}
    \left[\sin\theta\partial_\theta\left(\sin\theta
    \partial_\theta S\right)+\partial_\phi\partial_\phi S\right]
    +\frac{N\delta v^{\hat r}}{\sin\theta}\partial_\phi\partial_rS+
    \frac{1}{\sin\theta}\partial_t\partial_\phi S\right\}|_{r=R} \ ,
\end{equation}
 while the continuity condition for $E^{\hat \phi}$ (\ref{bound_e3}) 
gives a boundary condition for $\partial_\phi \Psi_{_{{\rm 
SC}}}|_{r=R}$:
\begin{equation}
\label{bound_Psi_3}
\Psi_{_{{\rm SC},\phi}}|_{r=R}=
    \left\{\frac{\delta v^{\hat\theta}}{R\sin\theta}
    \left[\sin\theta\partial_\theta\left(\sin\theta
    \partial_\theta S\right)+\partial_\phi\partial_\phi S\right]
    +N\delta v^{\hat r}\sin\theta \partial_\theta\partial_r S+ \sin\theta
    \partial_t\partial_\theta S\right\}|_{r=R} \ .
\end{equation}
 Integration of equation (\ref{bound_Psi_2}) or equation 
(\ref{bound_Psi_3}) over $ \theta $ or $ \phi $ respectively, gives a 
boundary condition for $\Psi_{_{\rm SC}}$. We will use the result of 
integrating equation (\ref{bound_Psi_2}) over $\theta$. Assuming that 
the perturbation depends on time $t$ as $e^{-i\omega t}$, we obtain 
the following condition, correct to first order in $\bar{\xi}$,
\begin{equation}
\label{bound_Psi}
\Psi_{_{{\rm SC}}}|_{r=R} =
    -\int \left\{\frac{\delta v^{\hat\phi}}{R\sin^2\theta}
    \left[\sin\theta\partial_\theta\left(\sin\theta
    \partial_\theta S_{0}\right)+
    \partial_\phi\partial_\phi S_{0}\right]
    +\frac{N\delta v^{\hat r}}{\sin\theta}
    \partial_\phi\partial_r S_{0}+
    \frac{1}{\sin\theta}
    \partial_t\partial_\phi (\delta S)\right\} d\theta|_{r=R}
    + e^{i\omega t} F(\phi) \ ,
\end{equation}
 where $F(\phi)$ is a function only of $\phi$ which we will determine 
below.

The components of the stellar-oscillation velocity field are 
continuously differentiable functions of $r, \, \theta $ and $\phi$. 
The boundary conditions for the electric field 
(\ref{bound_e2})-(\ref{bound_e3}) imply that the tangential components 
of the electric field $ \vec E_{_{\rm SC}} $ must be finite. The 
vacuum terms on the right-hand side of (\ref{ex_ef2})-(\ref{ex_ef3}) 
and the terms on both sides of equation (\ref{ex_ef3}) are also 
finite. Consequently, the term
\begin{equation}
\label{finite}
-\frac{\partial_\phi (\Psi_{_{\rm SC}})}{\sin\theta}|_{r=R}
\end{equation}
 should also be finite. Hence we obtain that $ \partial_\phi
(\Psi_{_{\rm SC}}) |_{\theta = 0, \pi; r = R } = 0 $ and so the 
function $F(\phi)$ in the expression for boundary condition
(\ref{bound_Psi}) must satisfy the condition $ (\Psi_{_{\rm SC}})
|_{\theta = 0, \pi; r = R } = C \, e^{ -i \omega\, t } $, where $ C $
is a constant. Using gauge invariance, we choose
\begin{equation}
\label{gauge}
\Psi_{_{\rm SC}}|_{\theta=0;r=R}=0 ,
\end{equation}
 and from this and equation (\ref{bound_Psi}), we obtain our 
expression for the boundary condition for $\Psi_{_{\rm SC}}$ at the 
stellar surface:
\begin{equation}
\label{bound_Psi_exact} \Psi_{_{{\rm
SC}}}|_{r=R}=-\int_{0}^{\theta} \left\{\frac{\delta
v^{\hat\phi}}{R\sin^2\theta}
    \left[\sin\theta\partial_\theta\left(\sin\theta
    \partial_\theta S_{0}\right)+
    \partial_\phi\partial_\phi S_{0}\right]
    +\frac{N\delta v^{\hat r}}{\sin\theta}
    \partial_\phi\partial_r S_{0}+
    \frac{1}{\sin\theta}\partial_t\partial_\phi (\delta S)\right\}
    d\theta|_{r=R}\ .
\end{equation}
%

\section{Toroidal oscillations of a NS with a dipole magnetic field}
\label{oscillation}

As an important application of this formalism, we now consider 
small-amplitude toroidal oscillations of a NS with a dipole magnetic 
field. For toroidal oscillations in the $ (\ell', m') $ mode, a 
generic conducting fluid element is displaced from its initial 
location $(r,\theta,\phi)$ to a perturbed location 
$(r,\theta+\xi^\theta,\phi+\xi^\phi)$ with the velocity field 
\citep{unno_89},
\begin{eqnarray}
\label{v_toroidal}
\delta v^{\hat r} = 0 \ , \qquad
\delta v^{\hat\theta} = \frac{d \xi^\theta}{d t} =
    e^{-i\omega t}\eta(r)\frac{1}{\sin\theta}\partial_\phi
    Y_{\ell' m'}(\theta ,\phi) \ , \qquad
\delta v^{\hat\phi} = \frac{d \xi^\phi}{d t} =
    -e^{-i\omega t}\eta(r)\partial_\theta
    Y_{\ell' m'}(\theta ,\phi) \ ,
\end{eqnarray}
 where $\omega$ is the oscillation frequency and $\eta(r)$ is the 
transverse velocity amplitude. Note that in the above expressions 
(\ref{v_toroidal}), the oscillation mode axis is directed along the 
$z$-axis. We use a prime to denote the spherical harmonic indices in
the case of the oscillation modes. 

\subsection{The unperturbed exterior dipole magnetic field}

If the static unperturbed magnetic field of the NS is of a dipole 
type, then the coefficients $s_{0\ell m}$ involved in specifying it 
have the following form (see eq. 117 of Paper I)
\begin{equation}
\label{slm} s_{0\,10}=-\frac{\sqrt{3\pi}}{2} \mu \cos\chi \ ,
\hskip 2.0 cm s_{0\,11}= \sqrt{\frac{3\pi}{2}} \mu \sin \chi\ ,
\end{equation}
where $ \mu $ is the  magnetic dipole moment of the star, as measured
by a distant observer, and $\chi$ is the inclination angle between the
dipole moment and $z$-axis. Substituting expressions (\ref{slm}) into
(\ref{sol_P1}) and then the latter into (\ref{exp_S0}), we get  
\begin{equation} \label{S_0}
S_{0}=-\frac{3 \mu  r^2}{ 8 M^3}\left[\ln N^2+
\frac{2M}{r}\left(1+\frac{M}{r}\right)\right](\cos\theta\cos\chi+e^{\rm
i \phi}\sin\theta\sin\chi)
\end{equation}
 The corresponding magnetic field components have the form
\begin{eqnarray}
\label{b0_1} && B_0^{\hat r} = - \frac{3 \mu }{4 M^3}
    \left[\ln N^2 + \frac{2M}{r}\left(1 +  \frac{M}{r}
    \right) \right] \; (\cos\chi\cos\theta +
    \sin\chi \sin\theta e^{{\rm i}\phi})
    \ ,
\\\nonumber\\
\label{b0_2} && B_0^{\hat \theta} = \frac{3 \mu N}{4 M^2 r}
    \left[\frac{r}{M}\ln N^2 +\frac{1}{N^2}+ 1
    \right] \; (\cos\chi \sin\theta
    - \sin\chi \cos\theta e^{{\rm i}\phi})
    \ ,
\\\nonumber\\
\label{b0_3} && B_0^{\hat \phi} = \frac{3 \mu N}{4 M^2 r}
    \left[\frac{r}{M}\ln N^2 +\frac{1}{N^2}+ 1
    \right]\; (- \rm i  \sin\chi e^{{\rm i}\phi})
    \ .
\end{eqnarray}
At the stellar surface, these expressions for the unperturbed 
magnetic field components become
\begin{eqnarray}
\label{mf_1} && B_{_R}^{\hat r} = f_{_R} B_{_0} (\cos\chi
\cos\theta +
    \sin\chi \sin\theta e^{{\rm i}\phi})
    \ , \qquad B_R^{\hat \theta} = h_{_R}B_{_0} (\cos\chi
\sin\theta
    - \sin\chi \cos\theta e^{{\rm i}\phi})
    \ , \qquad
 B_R^{\hat \phi} = -i h_{_R}B_{_0} (\sin\chi
e^{{\rm i}\phi})
    \ ,
\end{eqnarray}
where $ B_0 $ is defined as $ B_0 = 2 \mu / R^3 $. In Newtonian theory
$ B_0 $ would be the value of the magnetic strength at the magnetic
pole but this becomes modified in GR. The GR modifications are
contained within the parameters 
\begin{equation}
h_{_{R}}=\frac{3 R^2 N_{_R}}{8 M^2}
    \left[\frac{R}{M}\ln N_{_R}^2 +\frac{1}{N_{_R}^2}+ 1
    \right] \ , \quad  \quad
f_{_{R}}=-\frac{3R^3}{8M^3}\left[\ln N_{_R}^2+\frac{2M}{R}
    \left(1+\frac{M}{R}\right)\right]\ .
\end{equation}
For a given $ \mu $, the magnetic field near to the surface of the NS
is stronger in GR than in Newtonian theory, as already noted by
\citet{ginzburg_64}.

\subsection{The equation for $\Psi_{_{\rm SC}}$}
\label{section:main_results}

Substituting $S_0$ from (\ref{S_0}) into equation (\ref{Psi_0}), we 
obtain a partial differential equation containing two unknown 
functions $\Psi_{_{\rm SC}}$ and $\delta S$ for arbitrary oscillations 
of a NS with a dipole magnetic field
\begin{eqnarray}
\label{Psi_SC} -2r^2
q_1(r)\left(\cos\theta\cos\chi+e^{i\phi}\sin\theta\sin\chi\right)
\, \partial_r(\Psi_{_{\rm SC}})+\partial_r \left[r^2 q_1(r)\right]
\left(\sin\theta\cos\chi-e^{i\phi}\cos\theta\sin\chi\right)
\partial_{\theta}(\Psi_{_{\rm SC}})
\\\nonumber
-\partial_r \left[r^2 q_1(r)\right]\frac{e^{i\phi}\sin\chi}
{\sin\theta}\partial_{\phi}(\Psi_{_{\rm SC}})+ \frac{\partial_r
\left[r^2 q_1(r)\right]}{\sin\theta} \left[\left(\sin\theta\cos
\chi-e^{i\phi}\cos\theta\sin\chi\right)\partial_\phi\partial_t
(\delta S)+ie^{i\phi}\sin\chi\sin\theta\partial_\theta\partial_t
(\delta S)\right]=0\ ,
\end{eqnarray}
 where we have introduced a new function $ q_1(r) $ for simplicity of 
notation [see Eq. (\ref{short}) for the definition of $ q_1(r) $].

From (\ref{bound_Psi_exact}), the boundary condition for $ \Psi_{_{ 
\rm SC }} $ at the stellar surface is
\begin{equation}
\label{bound_Psi_dip}
\Psi_{_{{\rm SC}}}|_{r=R}=\int_{0}^{\theta} \left\{
 B_0 R f_{_{R}} \delta v^{\hat\phi}\partial_{\theta}\left(
\cos\theta\cos\chi+e^{i\phi}\sin\theta\sin\chi\right)
- \frac{1}{\sin\theta}\partial_t\partial_\phi
(\delta S)\right\} d\theta|_{r=R}\ .
\end{equation}
 Using the expressions for the velocity field of the toroidal 
oscillations (\ref{v_toroidal}) and for the boundary conditions for 
the partial derivatives of the SC potential 
(\ref{bound_Psi_2})-(\ref{bound_Psi_3}), we find that 
$\partial_t\delta S$ is given by (see Appendix A for details of the 
derivation)
\begin{eqnarray}
\label{ptdeltaS}
\partial_t\delta S(r,t)&=&\sum_{\ell=0}^{\infty}\sum_{m=-\ell}^{\ell}
\frac{B_{0}R f_{_{R}}\tilde{\eta}_{_{\rm
R}}}{\ell\left(\ell+1\right)} \frac{r^2\,q_{\ell}
\left(r\right)}{R^2\,q_{\ell} \left(R\right)}
\\\nonumber
&\times&\int_{4\pi}\left[\partial_{\theta} Y_{\ell
m}\left(\sin\theta\cos\chi-e^{i\phi}\cos\theta\sin\chi \right)
+ie^{i\phi}\partial_{\phi}Y_{\ell m}\sin\theta\sin\chi
\right]\frac{Y^*_{\ell' m'}(\theta,\phi)}{\sin\theta}d\Omega\ .
\end{eqnarray}
 From here on, for simplicity, we will consider only the case with $ 
\chi = 0 $. Although our solution depends on the angle between the 
magnetic field axis and the oscillation mode axis, focusing on the 
case $ \chi = 0 $ does not actually imply a loss of generality because 
any mode with its axis not aligned with a given direction can be 
represented as a sum of modes with axes along this direction. We have 
developed a MATHEMATICA code for analytically solving equation 
(\ref{Psi_0}) and hence obtaining analytic expressions for the 
electric and magnetic fields and for the SC density.

The solution of equation (\ref{Psi_SC}) for the case $ \chi = 0 $ is 
given in Appendix \ref{section:solution_for_chi_eq_0}, where we show 
that the general solution has the following form
\begin{equation}
\label{gen_sol_1} \Psi_{_{\rm SC}}=-\frac{1}{2}
    \frac{m'^{2}}{\ell'(\ell'+1)}
B_{0}R f_{_{R}}\tilde{\eta}_{_R}\, \int_{R}^{r}
\frac{\partial_{r'} [r'^2 q_1(r')]}{q_1(r')}\frac{q_{\ell'}(r')}
{R^2 q_{\ell'}(R)}\,\frac{Y_{\ell' m'}(\theta(r'),\phi)}
{\cos\theta (r')}\,dr' +\Phi_{2}\left[\sqrt{-r^2 q_1(r)}
\,\sin\theta,\phi,t\right]\ ,
\end{equation}
where $r'$ is the integration variable. In order to solve this
integral, the function $ \theta(r') $ is expressed in terms $ r' $ and
a constant $ \varphi_2 $ through the characteristic equation (\ref{char2})
and, after performing the integration, $ \varphi_2 $ is removed again
using (\ref{char2}). The unknown function $ \Phi_{2} $ is determined
using the boundary condition for $ \Phi_{2} |_{r=R}$ given by
(\ref{phi2}). Once the integral on the right-hand side of (\ref{phi2})
has been evaluated, we then  express all of the trigonometric
functions resulting from the integral, in terms of $ \sin \theta
$. This $ \theta $ is the value at $ r = R $. To obtain an
expression for the value of $ \Phi_2 $ at a general radius, we write
this $ \theta $ (at $ r = R $) in terms of the value of $ \theta $ at
a general point (with $ r > R $) using the characteristic relation
(\ref{char2}), i.e.,
\begin{equation}
\label{eqq:replacement_1}
\sin \theta \to \sqrt{ [ r^2 q_1 ( r ) ] / [ R^2 q_1 ( R ) ] }
\times \sin \theta \ ,
\end{equation}
so that
\begin{equation}
\label{eqq:replacement}
\cos \theta \to \left[1-\frac{r^2 q_1(r)}{R^2 q_1(R)} \sin^2
  \theta \right]^{1/2} \rm{  sign} (\cos \theta) \ , 
\end{equation}
where $ {\rm sign}(x) $ is defined such that $ {\rm sign}(x) = + 1 $ 
if $ x > 0 $, and $ {\rm sign}(x) = - 1 $ if $ x < 0 $. There are then
different expressions for $ \Psi_{_{\rm SC}} $ in the 
two regions $ \theta \in [0, \, \pi /2]$ and $ \theta \in [\pi/2, \, 
\pi]$. If these two expressions do not coincide at the equatorial 
plane for $ r > R $, then there will be a discontinuity in $\Psi_{_{\rm 
SC}} $ at $ \theta = \pi / 2 $, and quantities that depend on 
$\partial_\theta \Psi_{_{\rm SC}}$ will become singular there. As 
shown by TBS, the function $ \Psi_{_{\rm SC}} $ is indeed 
discontinuous at $\theta = \pi /2 $ for some oscillation modes and, as 
we discussed in Section \ref{lcda} above, this unphysical behaviour 
indicates that the LCDA ceases to be valid for those modes. In these 
cases, the accelerating electric field cannot be canceled without 
presence of strong currents which may become as large as 
(\ref{cur_density}) in some regions of the magnetosphere. The 
occurrence of such singularities was explained by TBS and the reader 
is referred to Section 3.2 of their paper for a detailed discussion.

Next we discuss how GR effects contribute to our solution. As
discussed above, for a given magnetic moment $ \mu $ (as measured by a
distant observer) the strength of the unperturbed magnetic field near
to the surface of the NS is larger in GR than in Newtonian theory. Due
to the linearity of the Maxwell equations, a perturbation of a stronger
magnetic field should produce a larger electric field for the same
oscillation parameters.  This in turn should lead to a larger absolute
value of the SC density in GR, since the SC density takes the value
necessary to cancel the electric field. In the next Section, we will
give a more quantitative analysis of the GR contribution in our
solution. 

We point out that the function $ \Psi_{_{ \rm SC }} $ does not depend
on $ \ell > 1 $ perturbations to the stellar magnetic field in the  
case of axisymmetric $ ( m' = 0 ) $ toroidal modes. This can be seen
from the fact that these perturbations are confined within the $\delta 
S$ terms which enter equation (\ref{Psi_SC}) for the function $
\Psi_{_{ \rm SC }} $ only through a derivative with respect to $ \phi
$; hence vanish for the axisymmetric modes. Therefore, the only
perturbation to the magnetic field is due to the $ \ell = 1 $ term,
and the solution for these modes is much simpler than that for
non-axisymmetric ($ m' \ne 0 $) modes. It is then convenient to
discuss separately the axisymmetric and non-axisymmetric cases.  

\begin{figure}
  \centerline{\epsfxsize = 8.0 cm
              \epsfbox{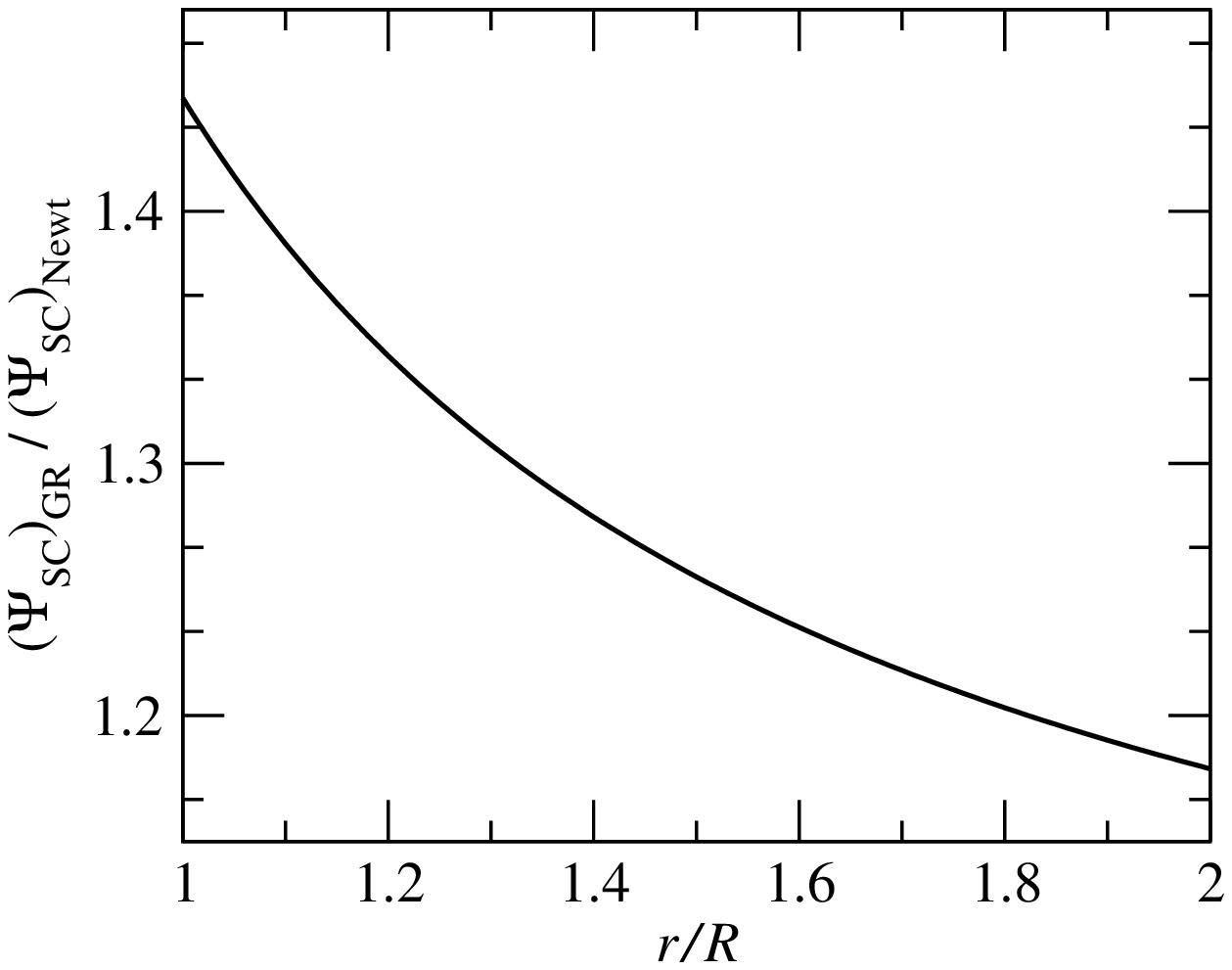}\qquad \quad
              \epsfxsize = 8.0 cm
              \epsfbox{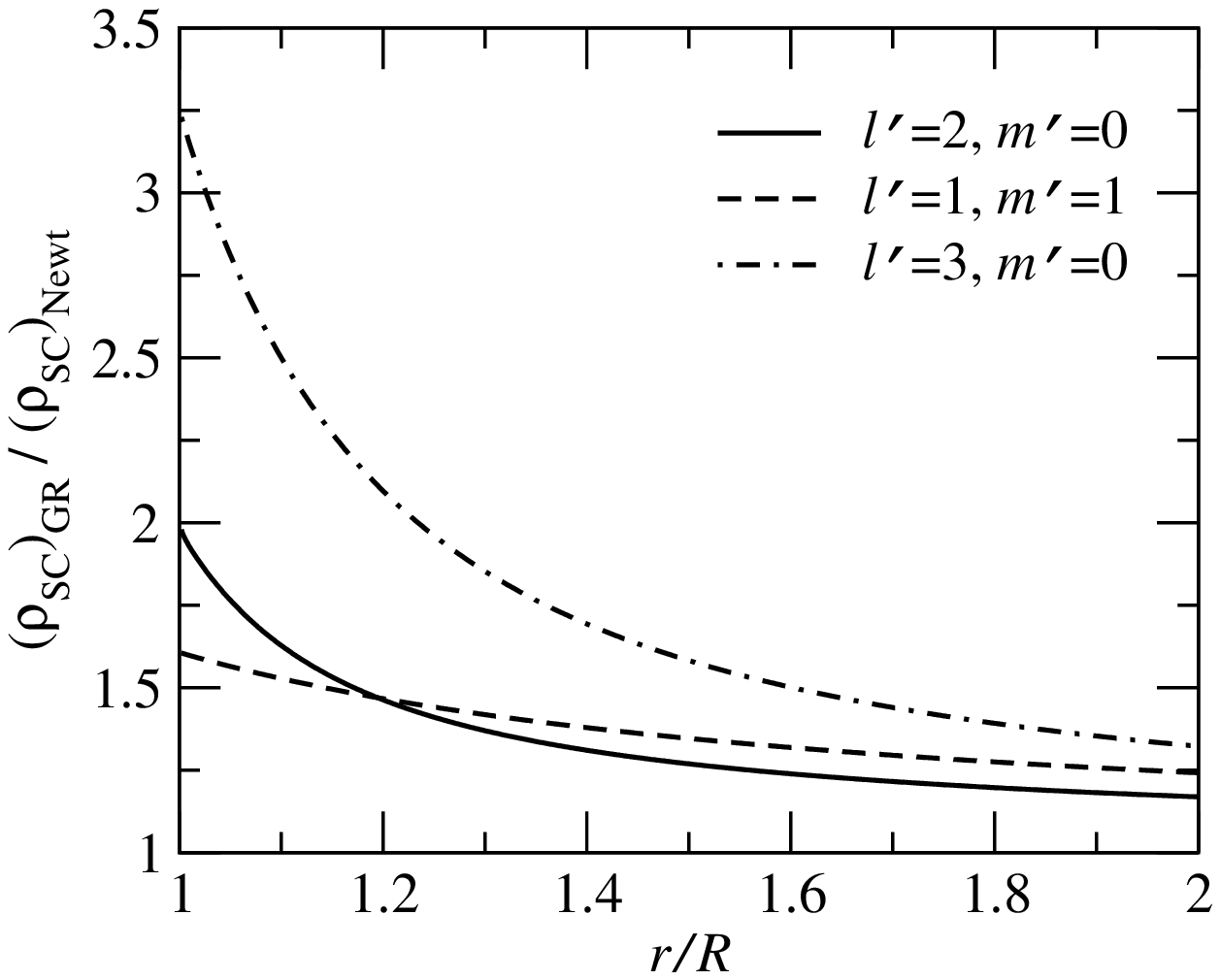}}
  \caption{\emph{Left panel:} The ratio $ ( \Psi_{_{ \rm SC }})_{_{
        \rm GR }} / ( \Psi_{_{ \rm SC }})_{_{ \rm Newt }} $ along the
    polar axis plotted as a function of the distance from the star,
    for axisymmetric toroidal modes of a NS with compactness $ M / R =
    0.2 $. \emph{Right panel:} The ratio $(\rho_{_{ \rm SC }})_{_{ \rm
        GR }} / (\rho_{_{ \rm SC }})_{_{ \rm Newt }}$ in the
    equatorial plane plotted as a function of $ r $, for a star with $
    M / R = 0.2 $, for toroidal oscillation modes $ (2,0) $, $ (1,1) $
    and $ (3,0) $.}
  \label{fig:ratio_psi_rho}
\end{figure}

The solution (\ref{gen_sol_1}) for case the $ m ' = 0 $ modes at $ r =
R $ has the following form
\begin{equation}
\label{eqq:psi_R} \Psi_{_{\rm
    SC}}\left(r,\theta,\phi,t\right)|_{r=R} = -B_{0} R f_{_{
    R}} \tilde{ \eta }_{_R}\int_{0}^{ \theta } \cos \vartheta \,
    \partial_{\vartheta} Y_{\ell' 0}(\vartheta,\phi) \ d\vartheta\ .
\end{equation}
 Using the properties of the spherical harmonics, we can express
$ \Psi_{_{\rm SC}} \left(r,\theta,\phi,t\right) |_{r=R} $ for
odd $ \ell' $ modes in the general form 
\begin{equation}
\label{eqq:psi_R_even_l} \Psi_{_{\rm SC}} \left(r, \theta, \phi, t
\right)|_{r=R} \sim \sum_{ n = 1 }^{{\cal N}} A_{2n} \sin^{2n} \theta
\ , 
\end{equation}
while for even $ \ell' $ modes, the general form of $ \Psi_{_{\rm SC}}
$ is
\begin{equation}
\label{eqq:psi_R_odd_l} \Psi_{_{\rm SC}} \left(r, \theta, \phi, t
\right)|_{r=R} \sim (A + B \cos \theta) \sum_{ n = 1 }^{{\cal N}}
A_{2n} \sin^{2n} \theta \ , 
\end{equation}
 where the coefficients $A$, $B$ and $A_{2n}$ do not depend on $ r $ 
and $ \theta $. The value of ${\cal N }$ equals $ \ell' / 2 + 1 $ for even $ 
\ell' $ and $ (\ell' + 1) / 2 $ for odd $ \ell' $. As we discussed
above, in order to obtain the solution for $ \Psi_{_{\rm SC}} $ for $
r > R $, one has use $ \sin\theta \to \sqrt{ r^2 q_1(r) / R^2 q_1(R) }
\ \sin\theta$ on the right-hand sides of (\ref{eqq:psi_R_even_l}) and
(\ref{eqq:psi_R_odd_l}). Thus the GR effects contribute to  
the solution for $ m' = 0 $ modes only through terms $ f_{_{ R}} 
\left[ r^2 q_1(r) / R^2 q_1(R)\right]^n $, where $ n \geq 1$. Note 
that the factor $ f_{_R} $ in this term appears due to the boundary
condition at the surface of the star, namely from the continuity of the 
tangential components of the electric field, while the factor $\left[ r^2 
q_1(r) / R^2 q_1(R)\right]^n $ appears due to the presence of
charged particles in the magnetosphere. The second factor is equal to $1$ at
the stellar surface and approaches its Newtonian value $ (R/r)^n $ at
large $r$ and small $ M / R $. Since $ f_{_R} > 1 $, the absolute value  
of $ \Psi_{_{ \rm SC }} $ should be greater in GR than in
Newtonian theory. For example, in the case of small $ \theta $, the
only term which is important is that with $ n = 1 $ and hence we
get $ ( \Psi_{_{ \rm SC }})_{_{ \rm GR }} / ( \Psi_{_{ \rm SC }})_{_{
    \rm Newt }} = f_{_{ R}} r^3 q_1(r) / R^3 q_1(R) $. This
quantity is shown in Figure~\ref{fig:ratio_psi_rho} (left panel),
where we can see that, near to the stellar surface, the function $
\Psi_{_{ \rm SC }} $ is larger in GR than in Newtonian theory, while
at larger $ r $ it asymptotically approaches its Newtonian value. 

\begin{figure}
  \includegraphics[width=165mm,angle=0]{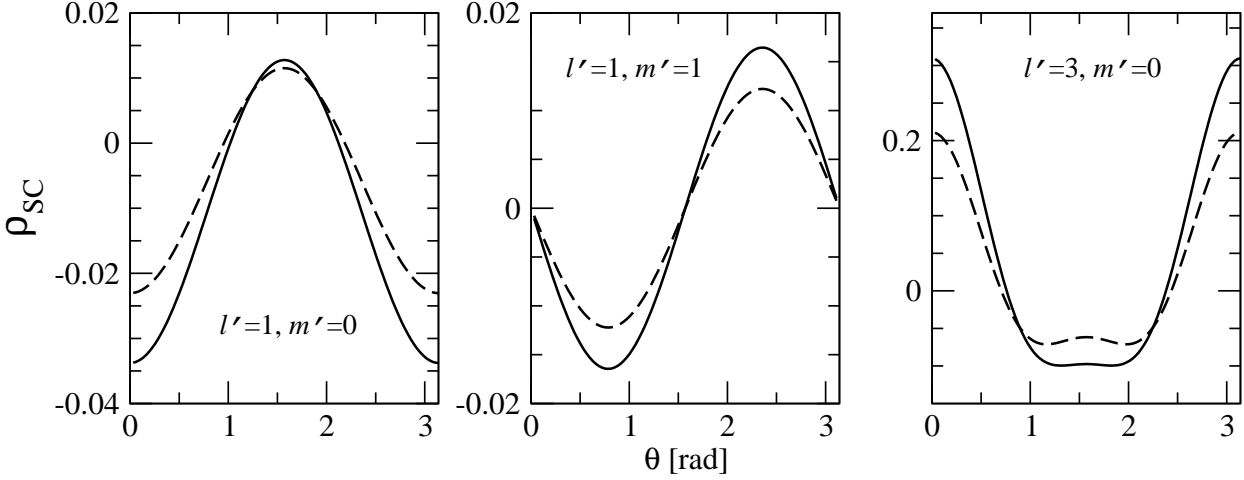}
  \caption{The space-charge density $ \rho_{_{\rm SC}} $ at $ r = 1.5
    R $ (in units with $ B_0 = \eta = 1 $) plotted as a function of
    $\theta$ for the toroidal oscillation modes $ (1,0) $, $ (1,1) $
    and $ (3,0) $. The solid lines refer to the SC density
    $\rho_{_{\rm SC}}$ in the relativistic theory, while its Newtonian
    value is shown by the dashed lines.}
  \label{fig:rho_vs_theta}
\end{figure}

Analysis of the GR contribution to the solution in the case of
non-axisymmetric modes is more complicated because in this case the 
solution depends not only on $ \ell = 1 $ perturbations to the 
magnetic field but also on $ \ell > 1 $ perturbations, which are
contained in the term $\partial_t \partial_\phi \delta S$ of equation
(\ref{Psi_SC}), and contribute to the solution due to the 
integral in (\ref{gen_sol_1}). Nevertheless, some rough estimates 
of the GR effects can be made in the following way. Near to the
stellar surface the integral in (\ref{gen_sol_1}) can be approximated
as  
\begin{equation}
\label{eqq:integral}
f_{_{ R}} \, \int_{R}^{r} \frac{\partial_z [z^2 q_1(z)]} {q_1(z)}
\frac{q_{\ell}(z)} {R^2 q_{\ell}(R)} \, \frac{Y_{\ell
    m}(\theta(z),\phi)} {\cos\theta (z)}\,dz \simeq f_{_{ R}} \,
\frac{r^2 q_{\ell}(r)} {R^2 q_{\ell}(R)} \,\frac{Y_{\ell m}
  (\theta(r),\phi)} {\cos\theta (r)} \ .
\end{equation}
 Close to the star, $ (r^2 q_{\ell}(r)) / (R^2 q_{\ell}(R)) \simeq 1 $ 
and so the leading GR contribution comes from the factor $ f_{_{ 
R}} $ which increases the absolute value of $\Psi_{_{\rm SC}}$ with 
respect to the Newtonian case. Further away from the star, $r^2 
q_{\ell}$ is approximately proportional to $ r^{-\ell} $ and so the 
integral in (\ref{eqq:integral}) can be approximated as $\sim f_{_R} 
(R/r)^{\ell'+1+m'/2}$ to leading order in $R/r$. Therefore, while this 
integral makes an important contribution to $\Psi_{_{\rm SC}}$ near to 
the star, it becomes negligibly small for $ \ell > 1 $ at $r > R$ as 
compared with $ \Phi_2 $. The GR effects contribute to $\Phi_2$ 
through the terms $ f_{_{ R}} \left[ r^2 q_1(r) / R^2 
q_1(R)\right]^n $ in a similar way to their contribution to 
$\Psi_{_{\rm SC }}$ for the axisymmetric modes discussed above. This
increase of the function $ \Psi_{_{\rm SC}} $ due to the GR effects
also lead to an increase in the absolute values of the SC density 
$\rho_{_{\rm SC}}$ near to the star, as shown in
Figures~\ref{fig:ratio_psi_rho} (right panel) and
\ref{fig:rho_vs_theta} for some toroidal oscillation modes.


\section{Energy losses}
\label{s_energy_loss}

It was shown by TBS that the kinetic energy of the stellar 
oscillations should be lost through being passed to plasma near to the 
stellar surface which then flows out along the open magnetic field 
lines. Note that within the framework of the TBS model, 
electromagnetic fields are considered only in the near zone and so the 
existence of the plasma outflow cannot be shown explicitly; however, 
qualitatively, the mechanism for the plasma outflow should be the 
following. The charged particles that were accelerated to high 
energies by the longitudinal electric field move along the magnetic 
field lines in the near zone. If the kinetic energy density of the
plasma at the equator becomes comparable to the energy density of the
magnetic field at some point, then the field line which crosses the
equator at that point becomes open. Plasma flowing along open field
lines forms an electromagnetically driven wind which closes at
infinity. There is then an electric current flowing along the stellar
surface between positive and negative emission regions. Because this
current must cross the magnetic field lines at the stellar surface, it
exerts a braking torque on the NS oscillations and thus reduces their
kinetic energy (see Section 3.2.2 of TBS for more details). 

In the following, we carry out a GR calculation of the energy lost by 
toroidal stellar oscillations due to plasma outflow. First we
calculate the energy losses due to the outflow of a particle along an
open field line from a given point on the stellar surface. For this
purpose, we start by considering the motion of the charged particle
along the $\theta$-direction on the stellar surface (where it crosses 
the magnetic field lines, hence exerts a braking torque on the stellar
oscillations). The equation of motion for a test particle of mass $m$
in a generic electromagnetic field has the general form \citep{landau_87} 
\begin{equation}
\label{eqq:eq_motion}
m\frac{D w^{\alpha}}{d \tau}=e F^{\alpha\beta}w_{\beta}\ ,
\end{equation}
where $ D/d \tau $ is a comoving derivative, $ w^{\alpha} $ is the
four-velocity of the particle given by
\begin{equation}
w_{\alpha} =\frac{u_{\alpha}+v_{\alpha}}{\sqrt{1-v^2}} \ ,
\end{equation}
$u^{\alpha}$ is the 4-velocity of the static observer (\ref{obs}), and
$ v^\alpha $ is the velocity of the particle relative to the static
observer. 

Because of time-invariance, there exists a timelike Killing vector 
$\xi^{\alpha}$ such that $\xi^{\alpha}\xi_{\alpha} = -N^2$. The 
four-velocity of the static observer can be expressed in terms of 
$\xi^{\alpha}$ as $u^{\alpha} = N^{-1}\xi^{\alpha}$, and therefore the 
energy of the particle is given by
\begin{equation}
{\cal E} = - p_{\alpha} \xi^{\alpha} = - m w_{\alpha} \xi^{\alpha}
\ . 
\end{equation}
Contracting the equation of motion (\ref{eqq:eq_motion}) with the
Killing vector $\xi_{\alpha}$ gives  
\begin{equation}
\label{eqq:eq_motion_2}
\xi_{\alpha}m w^{\alpha}_{;\beta}w^{\beta}=-e F^{\alpha\beta}
\frac{u_{\alpha}+v_{\alpha}}{\sqrt{1-v^2}}\xi_{\beta}\ ,
\end{equation}
The right-hand side of this can be rewritten as
\begin{equation}
-e F^{\alpha\beta}
\frac{u_{\alpha}+v_{\alpha}}{\sqrt{1-v^2}}\xi_{\beta} = e
F^{\alpha\beta}\frac{v_{\beta}\xi_{\alpha}} {\sqrt{1-v^2}} = 
e F^{\alpha\beta}N\frac{u_{\beta} v_{\alpha}} {\sqrt{1-v^2}} =
e \frac{E^{\hat \alpha} v_{\hat \alpha} N} {\sqrt{1-v^2}} \ ,
\end{equation}
while the left-hand side can be transformed as 
\begin{equation}
\label{eqq:relation_1}
\xi_{\alpha}m w^{\alpha}_{;\beta}w^{\beta} = m
(w_{\alpha}\xi^{\alpha})_{;\beta}w^{\beta} - m \xi_{\alpha;\beta}
w^{\alpha} w^{\beta} \ , 
\end{equation}
The second term on the right-hand of this equation vanishes due to
antisymmetry of the tensor $\xi_{\alpha;\beta}$. Therefore, the
projection of the equation of motion onto the Killing vector can be
written as
\begin{equation}
\label{eqq:eq_motion_3}
\frac{d {\cal E}}{d \tau} = e N \frac{ E^{ \hat \alpha } v_{ \hat
     \alpha }} {\sqrt{1 - v^2 }} \ .
\end{equation}
For particles moving along the $ \theta $-direction, this equation
takes the form 
\begin{equation}
d {\cal E} = e \frac{E^{\hat\alpha} R N_R^2 } {\sqrt{1-v^2}}
d\theta \ .
\end{equation}
Integrating this over the interval $(0,\theta)$, we get
\begin{equation}
\label{eqq:particle_energy}
\Delta {\cal E} = e R N_R^2
\int_0^{\theta} E^{\hat\alpha} d\theta .
\end{equation}
Note that in deriving this last equation, we have used the fact that $v
\ll 1 $. The quantity $ \Delta {\cal E}$ measures the energy that
would be carried away by a particle that leaves the surface of the star 
from a point with coordinates $(R,\theta,\phi)$. In order to calculate
the energy loss per unit time through a given surface element $ dS $,
$ \Delta {\cal E} $ needs to be multiplied by the the current density,
$ j^\alpha $ and integrated over the surface element $n_\alpha dS$,
where $n_\alpha$ is the unit spacelike vector orthogonal to the
surface $r = R$. The energy loss per unit time due to plasma emission
through the surface element $ dS $ is then given by
\begin{equation}
\label{eqq:total_power}
dL=\Delta {\cal E}(\theta,\phi) j^{\hat
\alpha} (R,\theta,\phi) n_{\hat \alpha}\ R^2\sin\theta d\theta
d\phi\ ,
\end{equation}
while the total energy loss $ L $ from the NS is obtained by
integrating $ d L $ over the entire open field line region on the
stellar surface. 

Now we determine the angle $ \theta_0 $ at which the last closed
magnetic field line intersects the stellar surface. Following TBS, we
define the last closed line as being that for which the kinetic energy
density of the outflowing plasma at the equator becomes 
equal to the corresponding energy density of the NS magnetic field. We
now derive a mathematical condition for this. The energy-momentum
tensor of the electromagnetic field is \citep{landau_87}
\begin{equation}
\label{energy_momentum_tensor}
T^{\alpha\beta}_{em} = \frac{1}{4\pi} \left( F^{\alpha\sigma}
F^{\beta}_{\sigma} - \frac{1}{4} g^{\alpha\beta} F^{\mu \nu} F_{\mu
  \nu} \right) \ .
\end{equation}
Using this expression, one can obtain an expression for the energy
density of the electromagnetic field
\begin{equation}
\label{particle_energy_2} {\cal
E}_{em}=NT^{\alpha\beta}u_{\alpha}u_{\beta} = N \frac{ B^2 + E^2
}{8 \pi }\ .
\end{equation}

The method for the calculation of the last closed field line proposed 
by TBS [equation (65) of TBS] is based on an implicit assumption that 
the plasma flows out from the star isotropically and its kinetic 
energy is distributed uniformly over the surface of a sphere with 
radius $ r = R_a $, where $ R_a $ is the radial coordinate of the 
point where the last closed field line crosses the equatorial plane. 
However, according to the definition of the last closed field line, 
the outflowing plasma should move along the field lines throughout the 
region $ r < R_a $ and so its energy cannot be distributed uniformly 
over the sphere $ r = R_a $. In the following, we derive an 
alternative formula for the calculation of $\theta_0$ that takes into 
account the correction due to the anisotropic plasma outflow along the 
magnetic field lines.

If the stellar magnetic field is dipolar, then the density of the field lines 
decreases with distance from the centre of the star. Since the charged 
particles move along the field lines, this implies that the kinetic energy 
density of a comoving element of plasma should decrease monotonically with $r$ 
as the field lines separate (we are assuming here that the plasma speed along 
the field lines is roughly constant). Since for dipole magnetic field lines $ 
f_r \sin^2 \theta / r = \mathrm{ const } $, it is easy to show the energy 
density of the plasma when it reaches the equatorial plane is smaller by a 
factor of $ 0.5 R^3/(R_a^3 f_{_R}) $ compared to its value at the stellar surface 
(assuming that the angle $ \theta $ is small). Therefore, the kinetic energy 
density of the outflowing plasma at the point $(R_a,0,\phi)$ is given by
\begin{equation} 
\label{eqq:epsilon_pl} 
\epsilon_{pl}(R_a,0,\phi) = \frac{1}{2f_{_R}} \frac{N_R} {N_{R_a}} 
\frac{R^3 }{R_a^3} \Delta {\cal E} j^{\hat \alpha} n_{\hat \alpha}\ , 
\end{equation} 
 where $N_{R_a}=\sqrt{1-2M/R_a}$ and the factor $ N_R / N_{R_a} $ 
accounts for the gravitational redshift of the energy of the plasma. 
The energy density of the magnetic field lines at the same point is 
determined from (\ref{particle_energy_2}) and has the following form
\begin{equation} 
\label{particle_energy_3} 
\epsilon_{em}(R_a,0,\phi) = \frac{N_{R_a}}{32\pi} \frac{R^6}{R_a^6} 
B_0^2 \ , 
\end{equation}
 Note that $R_a \gg R$ for small $\theta_0$, and so in this case 
$N_{_{\rm R_a }} \approx 1$ to good accuracy; from here on we will 
consider only the case of small $ \theta_0 $ and take $N_{_{\rm R_a}} 
= 1 $. Also $j^{\hat r} \gg j^{\hat \theta}$ for small $\theta$, and 
so in the following we will neglect the $\theta$ component of the 
current above the polar cap. The last closed field line is determined 
by the conditions
\begin{equation} 
\label{eqq:two_epsilons} 
\epsilon_{pl}(R_a,0,\phi) = \epsilon_{em}(R_a,0,\phi)\ . 
\end{equation} 
 Substituting (\ref{particle_energy_3}) and (\ref{eqq:epsilon_pl}) 
into (\ref{eqq:two_epsilons}), one can obtain an algebraic equation
\begin{equation} 
\label{eqq:theta_0} 
16 \pi N_R \Delta {\cal E} \rho_{_{\rm SC}} = f_{_R}^4 B_0^2 \theta_0^6 \ 
\end{equation} 
 for determining $\theta_0$ after expressing $ \Delta {\cal E} $ in terms of $ 
\theta_0 $. Once we know $\theta_0$, we can calculate the total energy of the 
outflowing plasma by integrating (\ref{eqq:total_power}) over the entire open 
field line region of the stellar surface:
\begin{equation} 
\label{eqq:en_loss} 
L = \int_0^{2\pi}d\phi \int_0^{\theta_0} d\theta\ |j^{\hat 
r}(R,\theta,\phi) \ \Delta {\cal E} (\theta,\phi)| R^2 \sin\theta\ . 
\end{equation} 
 The motion of charged particles in strong magnetic fields can be approximated 
as a relativistic motion along the field lines. This is a reasonable 
approximation because the electrons are estimated to become relativistic at a 
height of a few centimeters above the stellar surface. Therefore in the frame 
of the static observer we take \citep{muslimov_92}
\begin{equation} 
\label{eqq:net_cur} 
j^{\hat r}=\rho c\frac{B^{\hat r}}{B} 
\end{equation} 
 where $\rho(R,\theta,\phi) $ is the SC density.

TBS solved the Newtonian version of (\ref{eqq:theta_0})\footnote{It
  should be noted here that, as we mentioned above, TBS uses a 
  slightly different formula for the determination of $\theta_0$ which 
  does not take into account the anisotropy of the plasma outflow.} 
and calculated $L$ for three modes: (1,1), (2,0) and (3,0). They also 
estimated the order of magnitude of the energy loss $L$ and the angle 
$\theta_0$ for all of the oscillation modes for which $\theta_0$ is 
small. They found that $\theta_0$ is small for modes with $m'<3$ and, 
as we will see later, this result also holds in GR. We will next solve 
the relativistic equation (\ref{eqq:theta_0}) and calculate $\theta_0$ 
and $L$ for {\it all} of the modes $(\ell',m')$ for which the angle 
$\theta_0$ is small. We will present elsewhere a study of the case for 
large $\theta_0$.

In order to solve equation (\ref{eqq:theta_0}) and calculate $L$ in 
the case of small $\theta_0$, we use the following procedure. First, we 
approximate equation (\ref{Psi_SC_fin_m}) and the boundary condition 
(\ref{bound_tor}) for small $\theta$ by expanding all of the 
functional dependence on $\theta$ in a Taylor series, taking into 
account only terms of the two lowest orders in $ \theta $. Secondly, 
we solve this equation and calculate $\rho_{_{\rm SC}}$ and $\Delta 
{\cal E}$ using the approximate solution for $\Psi_{_{\rm SC}}$. Then, 
we calculate the angle $ \theta_0 $ by substituting $ \rho_{_{\rm SC}} 
$ and $ \Delta {\cal E} $ into (\ref{eqq:theta_0}). Finally, 
substituting the expressions for $\theta_0$, $\rho_{_{\rm SC}}$ and $ 
\Delta {\cal E} $ into (\ref{eqq:en_loss}), gives us the expression 
for $L$. Since the methods for solving equation (\ref{Psi_SC_fin_m}) 
for $ m' = 0 $ and $ m' \ne 0 $ exploit a similar technique, we 
present just the solution for $ m' = 0 $ below, in this Section, while 
the solution for non-axisymmetric modes is presented in Appendix 
\ref{section:energy_loss_m_ne_0}.

For small angles $\theta$, the spherical harmonic $ Y_{\ell 
m}(\theta,\phi)$ can be approximated by the sum of the first two 
lowest order terms in the expansion in terms of $\theta$
\begin{equation}
\label{expan_sh} Y_{\ell m}(\theta,\phi)\approx A_{\ell
m}^{(1)}(\phi)\,\theta^m+A_{\ell m}^{(2)}(\phi)\,\theta^{m+2}\ .
\end{equation}
 In this case, the characteristics (\ref{char2}) take the form
\begin{equation}
\varphi=\sqrt{-r^2 q_1(r)}\,\theta\ .
\end{equation}
 Specialising to small $ \theta $ in the boundary condition 
(\ref{bound_tor}) for $\Psi_{_{\rm SC}}$, gives the following 
expression:
\begin{equation}
\label{bc_app} \Psi_{_{\rm SC}}|_{r=R}=-B_{0} R f_{_{ R}}
\tilde{\eta}_{_R} A_{\ell' 0}^{(2)}\,\theta^2 \ .
\end{equation}
 In order to calculate $\Psi{_{\rm SC}}$ for arbitrary $r$, one has to replace 
$\theta$ in this equation by $\sqrt{[r^2 q_1(r)]/[R^2 q_1(R)]}\,\theta$, 
following the argument given earlier for $\sin\theta$, and this then gives the 
solution:
\begin{equation}
\label{psi_app} \Psi_{_{\rm SC}}=-B_{0} R f_{_{
R}}\tilde{\eta}_{_R} \,\frac{r^2 q_1(r)}{R^2 q_1(R)}A_{\ell'
0}^{(2)}\,\theta^2\ .
\end{equation}
 Substituting this into formula (\ref{SC_charge}), we obtain
the expression for the SC density:
\begin{equation}
\label{SC_chrge_app} \rho_{_{\rm SC}}=\frac{B_{0} f_{_{ R}}
\tilde{\eta}_{_R} }{\pi\, N_{_{\rm R}}R}\frac{r^2 q_1(r)}{R^2
q_1(R)}A_{\ell' 0}^{(2)}\ .
\end{equation}
 Substituting $\Psi{_{\rm SC}}$ as given by (\ref{psi_app}) into 
(\ref{ex_ef2}), we obtain the expression for electric field $E_{_{\rm 
SC}}^{\hat \theta}$. Substituting this into 
(\ref{eqq:particle_energy}) gives $ \Delta {\cal E} $:
\begin{equation}
\label{pot_diff_app} \Delta {\cal E}(\theta,\phi) = B_{0} R
f_{_{ R}}\tilde{\eta}_{_R} N_{_{\rm R}}A_{\ell' 0}^{(2)}\,
\theta^2\ .
\end{equation}
 Using this expression for $\Delta {\cal E}$ in equation 
(\ref{eqq:theta_0}), we obtain an algebraic equation for $\theta_0$, 
which has the following solution:
\begin{equation}
\label{eqq:theta_0_m0} \theta_0 = 2 N_R^{1/4} \left[ \frac{ \tilde
    \eta_{_R} A_{\ell' 0}^{(2)} }{ f_{_{ R}} } \right]^{1/2} \ .
\end{equation}
 Substituting (\ref{pot_diff_app}) and (\ref{eqq:net_cur}) into 
(\ref{eqq:en_loss}), gives us the following expression for the total energy 
loss:
\begin{equation}
\label{en_loss_tor} L_{\ell' 0}=\frac{1}{2} \left[ B_{0} R f_{_{
      R}} \tilde \eta_{_R} A_{\ell' 0}^{(2)} \right]^2 \theta_0^4\ ,
\end{equation}
 and then using (\ref{eqq:theta_0_m0}) in equation 
(\ref{en_loss_tor}), gives
\begin{equation}
\label{en_loss_tor2}  L_{\ell' 0} = 8 N_R \left[ B_{0} R \ \tilde
  \eta_{_R}^2 (A_{\ell' 0}^{(2)})^2 \right]^2 \ .
\end{equation}
 The energy losses for non-axisymmetric modes are calculated in Appendix 
\ref{section:energy_loss_m_ne_0}. The angle $ \theta_0 $ and the energy loss $ 
L $ are given by expressions (\ref{eqq:theta_0_mne0}) and 
(\ref{eqq:energy_loss_mne0}) in that case. Note that determining $\theta_0$ 
using the direct relativistic extension of the TBS method would give a value 
that is smaller than (\ref{eqq:theta_0_m0}) by a constant factor of $ 2^{0.5} 
$, causing $L$ to be smaller by a factor of $4$. For the $ m' = 1 $ modes, 
these correction factors are $ (2 \pi)^{0.25} $ and $ 2 \pi $ respectively, 
while for the $ m' = 2 $ modes they are $ (3 \pi)^{0.5} $ and $ (3 \pi)^3 $.

\begin{figure}
  \includegraphics[width=165mm,angle=0]{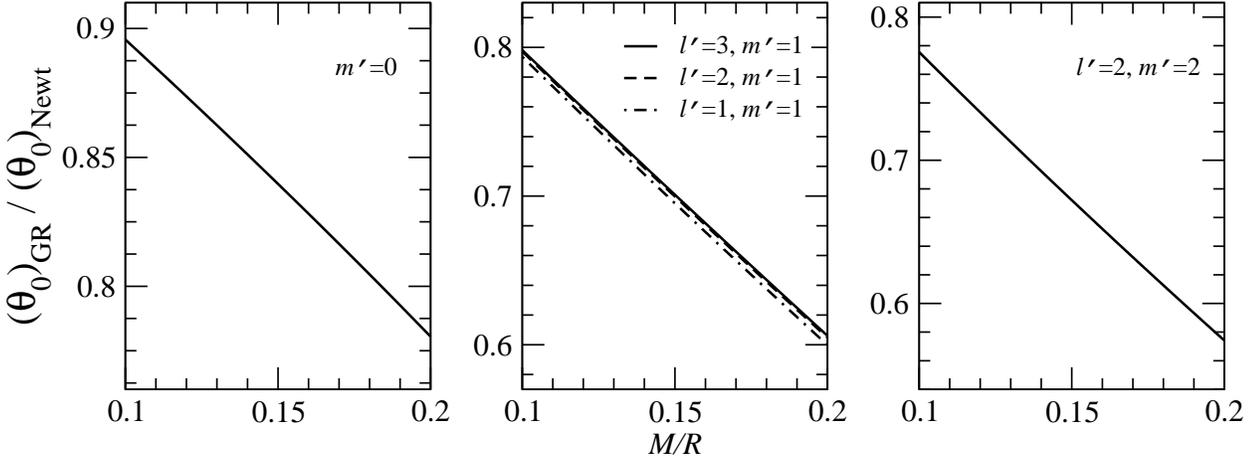}
  \caption{The ratio $(\theta_0)_{_{\rm GR}}/(\theta_0)_{_{\rm Newt}}$
  is plotted against stellar compactness $ M / R $ for some
  representative toroidal oscillation modes with $ m' = 0, \ 1, \ 2 $;
  it asymptotically approaches 1, of course, in the Newtonian limit $
  M / R \to 0$.} 
  \label{fig:ratio_theta0}
\end{figure}

We now continue our discussion with analysis of the contribution of
the effects of GR to the quantities $ \theta_0$ and $L_{\ell' 0}$. The
ratio of $ \theta_0 $ in GR to its Newtonian counterpart for a given
oscillation amplitude [i.e., for $ ( \tilde  \eta_{_R} )_{_{ \rm GR }
  } = ( \tilde \eta_{_R} )_{_{ \rm Newt }} $] can be obtained using 
equation (\ref{eqq:theta_0_m0}):
\begin{equation}
\label{ratio_theta_0}
\frac{\left(\theta_0\right)_{_{\rm GR}}}{\left( \theta_0 \right)_{_{
  \rm Newt}}} = \frac{N_R^{1/4}} {f_{_{ R}}^{1/2} } \ .
\end{equation}
 This ratio is shown in Figure~\ref{fig:ratio_theta0} (left panel) plotted as a 
function of the stellar compactness parameter $ M / R $. It can be seen that in 
GR, the angle $ \theta_0 $ is smaller than in Newtonian theory (it approaches 
its Newtonian value as $ M / R \rightarrow 0 $), meaning that GR effects lead 
to the polar cap being smaller. The reason for this can be understood in the 
following way. The angle $ \theta_0 $ is determined from equation 
(\ref{eqq:theta_0}) and it is convenient to analyze the relativistic effects by 
looking at this equation. The contribution from the curvature is contained here 
in three terms: $i$) the factor $f_{_R}^4$ on the right-hand side, $ii$) the 
factor $N_R $ on the left-hand side and $iii$) a factor $ f_{_R}^2 $ on the 
left-hand side that is contained implicitly within the term $\Delta \cal{E} 
\rho_{_{\rm SC}}$. The first of these three terms accounts for the modification 
of the geometry of the background dipole magnetic field lines due to the 
curvature. As we discussed above, the magnetic field lines obey the relation $ 
f_r \sin \theta / r^2 = \mathrm{const} $, with $ f_r = 1 $ in the Newtonian 
case but $f_r > 1$ in GR. This means that a magnetic field line with a given 
energy density at $ r = R_a $ on the equatorial plane, crosses the stellar 
surface closer to the pole in GR than in Newtonian theory. The second 
relativistic factor accounts for the redshift of the kinetic energy of the 
plasma due to the displacement from $ r = R $ to $ r = R_a$. Finally, the third 
factor is responsible for the amplification of the energy density of the 
outflowing plasma caused by the increase of the magnetic field strength (for a 
given magnetic moment) due to GR effects. The last of these would obviously 
lead to an increase of $\theta_0$ in GR if it were acting alone. However, this 
effect is counteracted by the change of the geometry of the magnetic field 
lines and the gravitational redshift and those two effects are substantially 
stronger, giving the result that the angle $\theta_0$ is smaller in GR than in 
Newtonian theory. Note that, as expected, the ratio $ \left( \theta_0 
\right)_{_{ \rm GR }} / \left( \theta_0 \right)_{_{ \rm Newt}}$ for the 
axisymmetric toroidal modes does not depend on $\ell'$, while for the general 
$ m' \ne 0 $ modes, there {\em is} dependence on $\ell'$. This is because, as 
discussed above, the function $ \Psi_{_{\rm SC}} $ for $ m'=0$ depends only on 
the lowest $ \ell = 1 $ perturbation to the magnetic field since the higher $ 
\ell $ perturbations are contained in the term $ \partial_\phi \partial_t 
\delta S$ which disappears in axisymmetry.

The ratio of the angle $ \theta_0 $ in GR to its Newtonian 
counterpart, for $ m' \ne 0$ modes, can be obtained using equation 
(\ref{eqq:theta_0_mne0}) (see Appendix 
\ref{section:energy_loss_m_ne_0} for details of the derivation):
\begin{equation}
\label{eqq:ration_theta_0_mne0}
 \frac{\left( \theta_0\right)_{_{\rm GR}}}{\left( \theta_0 \right)_{_{
  \rm Newt}}}  =  \left( \frac{N_R}{f_{_R}^2} \frac{| D_{\ell'
  m'}|_{_{\rm GR}} }{| D_{\ell' m'}|_{_{\rm Newt}} } \right)^{
  \frac{1}{6 - 2 m'} } \ .
\end{equation}
 Figure~\ref{fig:ratio_theta0} (central and right panels) shows the 
ratio of angle $\theta_0$ in GR to its Newtonian equivalent for the 
modes $ m' = 1, 2 $ for several values of $ \ell' $. As can be seen in 
these plots, $ \theta_0 $ is smaller in GR than in Newtonian theory 
also for the $ m' = 1 $ and $ m' = 2 $ modes. The reason for this is 
similar to that for the $m'=0$ modes which we discussed above. We have 
made the analysis for values of $ \ell' $ up to $ \ell' = 10 $ and the 
ratio $ \left( \theta_0\right)_{_{\rm GR}} / \left( \theta_0 
\right)_{_{ \rm Newt}} $ was found to be rather insensitive to the 
values of $ \ell' $. (The dependence on $ \ell' $ is due to the fact 
that $\left( \rho_{_{\rm SC}} \right)_{_{\rm GR}} / \left( \rho_{_{\rm 
SC}} \right)_{_{\rm Newt}}$ is generally larger for higher $\ell'$ at 
small $ \theta_0 $.)

\begin{figure}
  \includegraphics[width=165mm,angle=0]{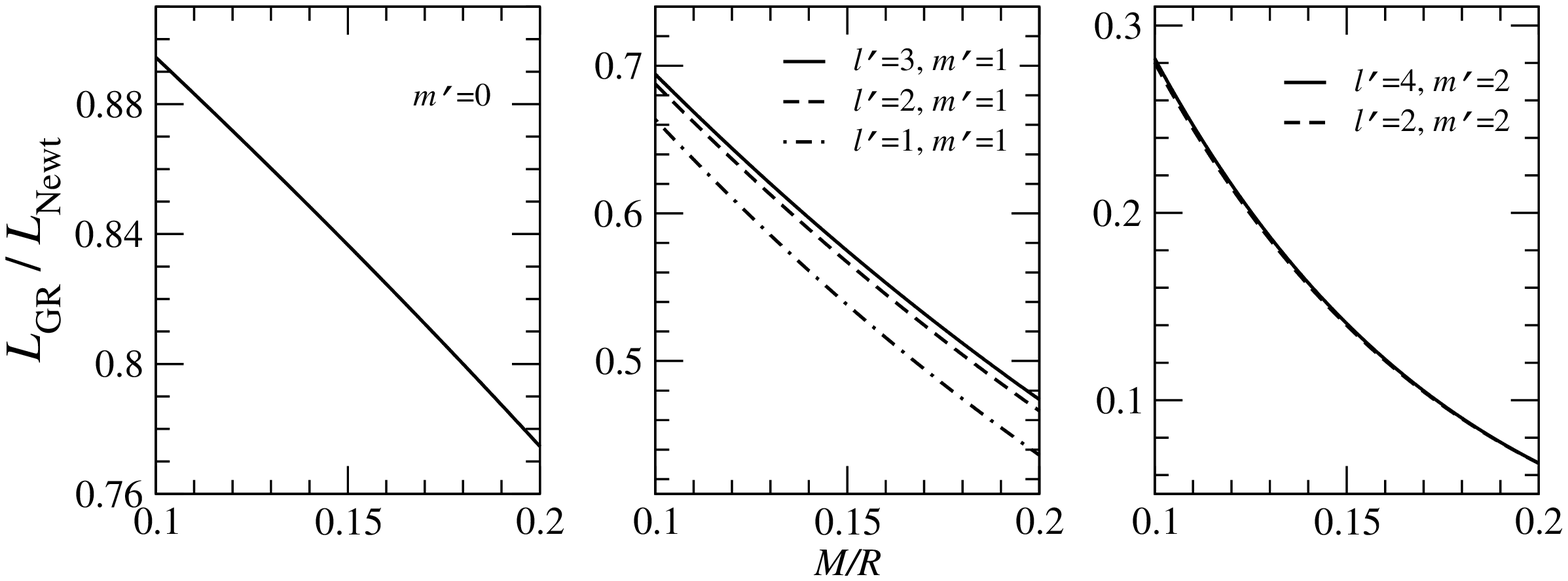}
  \caption{The ratio $L_{_{\rm GR}}/L_{_{\rm Newt}}$ is plotted
    against stellar compactness $ M / R $ for toroidal oscillation
    modes with $ m' = 0, \ 1, \ 2 $.}
  \label{fig:ratio_L}
\end{figure}

The ratio of the energy losses for the $ m' = 0 $ modes in GR and in Newtonian 
theory is equal to $N_R$, as one can obtain using equation 
(\ref{en_loss_tor2}), and this is plotted in the left panel of Figure 
\ref{fig:ratio_L}. The GR modification is caused by the gravitational redshift of 
the plasma energy; the other ``magnetic'' GR effects do not influence this, 
because the shrinking of the polar cap and the increase in the plasma energy 
density exactly compensate each other.

The ratio of the energy losses for the $ m' \ne 0 $ modes in GR and in 
Newtonian theory can be straightforwardly obtained using equation 
(\ref{eqq:energy_loss_mne0}):
\begin{equation}
  \label{eqq:ration_l_mne0}
 \frac{\left( L_{\ell' m'}\right)_{_{\rm GR}}}{\left( L_{\ell' m'}
  \right)_{_{ \rm Newt}}} =  f_{_R}^4 N_R^{\frac{1+m'}{3-m'}}
 \left( f_{_R}^{-2} \frac{|
  D_{\ell'  m'}|_{_{\rm GR}} }{| D_{\ell' m'}|_{_{\rm Newt}} }
  \right)^{ \frac{4}{3-m'} } \ .
\end{equation}
 This is also shown in Figure~\ref{fig:ratio_L} where one can see that the 
energy loss of the $ m' \ne 0 $ modes is smaller in GR than in Newtonian 
theory. The reason for this is the same as for the $ m' = 0 $ modes discussed 
above, i.e., the increase in the energy density of the outflowing plasma cannot 
compensate the shrinking of the polar cap. Moreover, for the $ m' = 2 $ modes, 
the energy density of the outflowing plasma is proportional to $ \theta^{2m'}$ 
(see eqs. (\ref{eqq:particle_energy}) and (\ref{eqq:rho_sc_small_theta})) and 
so the total energy losses are very sensitive to the size of the polar cap. 
Because of this, the ratios of the total energy losses for the $ m' = 2 $ modes 
in GR and in Newtonian theory are much smaller than those for the $ m' = 0 $ 
and $ m' = 1 $ modes.
%

\section{Summary}

In this paper, we have described our general relativistic model for the 
force-free magnetosphere of an oscillating, non-rotating neutron star. Our 
approach is based on the previous Newtonian model developed by TBS and focuses 
on toroidal modes which are thought to be particularly relevant for magnetar 
QPO phenomena. We have taken the spacetime geometry to be spherically symmetric 
and have neglected any modifications of it caused by the electromagnetic 
fields, the stellar oscillations and the magnetospheric plasma. Within this 
context, we have derived the general relativistic Maxwell equations for 
arbitrary small-amplitude oscillations of a neutron star with a generic 
magnetic field configuration and have shown that, as in the Newtonian case, 
they can be solved analytically for the force-free configuration of the 
electromagnetic fields (i.e. $E_{||} \ll B $ ) under the low current density 
approximation (LCDA). We have applied our formalism to small-amplitude toroidal 
oscillations of a neutron star with a dipole magnetic field and have found that 
the LCDA is valid for at least half of these modes in GR, as in the Newtonian 
calculations of TBS. We have also discussed the contribution of the GR effects 
to our solution, finding that they lead to an increase in the absolute values 
of the electromagnetic fields and the space charge (SC) density near to the 
stellar surface.

We have calculated the energy losses due to plasma outflow resulting from these 
oscillations, focusing on cases where the size of the polar cap is small so 
that one can expand the Maxwell equations as Taylor series in powers of $ 
\theta $, retaining only the two lowest order terms. This approach leads to a 
great simplification and allowed us to perform a thorough analysis of the 
solution. We have found that in GR, the polar cap is smaller than in Newtonian 
theory and have shown that this is due to the change in the geometry of the 
dipole magnetic field and the gravitational redshift of the energy of the 
outflowing plasma. Also, we found that the oscillation modes which have small $ 
\theta_0 $, all have $ m' < 3 $ as in the Newtonian case.

The total energy loss resulting from the stellar oscillations causing plasma 
outflow through the polar cap region, is determined through an integral over 
the whole polar cap area, and so it depends on both the kinetic energy density 
of the outflowing plasma and the surface area of the polar cap. Although GR 
effects lead to some increase in the energy density of the outflowing plasma 
(due to the increase in the surface magnetic field strength for a given 
magnetic moment), the area of the polar cap is smaller in GR and we have found 
that the increase in the energy density of the outflowing plasma cannot 
compensate for the shrinking in size of the polar cap. Therefore the total 
energy losses for the toroidal oscillation modes are significantly smaller in GR 
than in Newtonian theory.

In conclusion, we point out that while our calculations represent an advance 
with respect to previous ones, they still do not include a number of very 
important aspects which would be necessary for a realistic description of these 
phenomena. Most importantly, as noted above, they do not take account of 
electric currents flowing in the magnetosphere. Inclusion of these in a 
consistent way would require solving a version of the non-linear ``pulsar 
equation'' \citep{michel_73} for oscillating neutron stars, possibly by 
adopting a numerical approach similar to that of \citet{contopoulos_99} 
\citep[see also][]{gruzinov_05, timokhin_06} or performing {\it time-dependent} 
  simulations of the magnetosphere \citep{spitkovsky_06, komissarov_06,
  mckinney_06}. This will be the subject of future investigations.

\section*{Acknowledgments}

We wish to thank Luciano Rezzolla for bringing this problem to our attention 
and for many helpful comments. We also acknowledge helpful discussions with V.  
S. Beskin, U. Geppert, K. Glampedakis, A.V. Khugaev and
A.N. Timokhin. BJA gratefully acknowledges travel support from the
Exchange of Astronomers Programme of the IAU C46-PG-EA and the ICTP
Associate Programme.

\appendix

\section{Calculation of $\delta S$}

The derivative with respect to $\phi$ of the right hand side of 
equation (\ref{bound_Psi_2}) must be equal to the derivative with 
respect to $\theta$ of the left-hand side of (\ref{bound_Psi_3}). 
Using this condition and expression (\ref{exp_P2}) for $ S $, we 
obtain
\begin{eqnarray}
\label{delta_p_lm_2} \triangle_{\Omega} (\partial_{t} \delta
S)|_{r=R} = & - & \Bigg[\left( N \delta v^{\hat r}\partial_{r}
\triangle_{\Omega}S_{0}+\frac{\delta v^{\hat\theta}}{r}
\partial_{\theta}
\triangle_{\Omega}S_{0}+\frac{\delta v^{\hat\phi}}{r\sin\theta}
\partial_{\phi}\triangle_{\Omega}S_{0}\right)
\nonumber \\\nonumber \\
& + &
\frac{\triangle_{\Omega}S_{0}}{r\sin\theta}
\left(\partial_{\theta}(\sin\theta \delta
v^{\hat\theta})+\partial_{\phi}\delta v^{\hat\phi}\right)
 + N\partial_{r}\partial_{\theta} S_0\,\partial_{\theta}\delta
v^{\hat r}+ \frac{N}{\sin^2\theta}\partial_{r}\partial_{\phi}S_0
\,\partial_{\phi}\delta v^{\hat r}\Bigg]|_{r=R}. \
\end{eqnarray}
 We now introduce a new function for shorthand:
\begin{equation}
\label{short}
q_{\ell}(r)=-M\frac{d}{dr}\left[r\left(1-\frac{2M}{r}\right)\frac{d}{dr}
Q_{\ell}\left(1-\frac{r}{M}\right)\right]\ .
\end{equation}
 Using this notation, one can rewrite equations (\ref{sol_P2}) and
(\ref{exp_dS}) in the following way
\begin{eqnarray}
\label{exp_dSlm} &&\delta S_{\ell m}(r,t)=\frac{r^2
q_{\ell}(r)}{M^3}\,\delta s_{\ell m}(t), \\
&&\label{exp_dS2}\delta
S(t,r,\theta,\phi)=\sum_{\ell=0}^{\infty}\sum_{m=-\ell}^{\ell}
\frac{r^2 q_{\ell}(r)}{M^3}\,\delta s_{\ell m}(t)Y_{\ell
m}(\theta,\phi).
\end{eqnarray}
 Substituting the right hand side of the last equation into the left
hand side of (\ref{delta_p_lm_2}), we get
\begin{eqnarray}
\label{deltaplm3}
\frac{r^2}{M^3}\sum_{\ell=0}^{\infty}\sum_{m=-\ell}^{\ell}\,
q_{\ell}\left(r\right)\,\partial_t\delta s_{\ell
  m}(t)\,\ell(\ell+1)Y_{\ell m}|_{r=R} &=& \Bigg[\left( N
\delta v^{\hat r}\partial_{r} \triangle_{\Omega}S_{0}+\frac{\delta
v^{\hat\theta}}{r}
\partial_{\theta} \triangle_{\Omega}S_{0}+\frac{\delta
v^{\hat\phi}}{r\sin\theta}\partial_{\phi}
\triangle_{\Omega}S_{0}\right)
\\\nonumber\\\nonumber
+\,\, \frac{\triangle_{\Omega}S_{0}}{r\sin\theta}
\left(\partial_{\theta}(\sin\theta \delta v^{\hat\theta})
+\partial_{\phi}\delta v^{\hat\phi}\right)
&+&N\partial_{r}\partial_{\theta} S_0\,\partial_{\theta}\delta
v^{\hat r}+ \frac{N}{\sin^2\theta}\partial_{r}\partial_{\phi}S_0
\,\partial_{\phi}\delta v^{\hat r}\Bigg]|_{r=R}.
\end{eqnarray}
 Multiplying now both sides of this equation by $Y_{\ell' 
m'}^{*}(\theta,\phi)$, integrating it over solid angle $4\pi$ and 
using the orthogonality condition for spherical harmonics $\int 
Y_{\ell' m' }^{*}Y_{\ell m}d\Omega=\delta_{\ell'\ell}\delta_{m' m}$, 
we obtain
\begin{eqnarray}
\label{deltaplm4} \partial_t\delta s_{\ell m}(t) &=&
\frac{M^3}{\ell'(\ell'+1)\,r^2q_{\ell'}\left(r\right)}\,\int_{4
\pi} \Bigg[\left( N \delta v^{\hat r}\partial_{r}
\triangle_{\Omega}S_{0}+\frac{\delta v^{\hat\theta}}{r}
\partial_{\theta} \triangle_{\Omega}S_{0}+\frac{\delta
v^{\hat\phi}}{r\sin\theta}\partial_{\phi}
\triangle_{\Omega}S_{0}\right)
\\\nonumber\\\nonumber
&+& \frac{\triangle_{\Omega}S_{0}}{r\sin\theta}
\left(\partial_{\theta}(\sin\theta \delta v^{\hat\theta})
+\partial_{\phi}\delta v^{\hat\phi}\right)
+N\partial_{r}\partial_{\theta} S_0\,\partial_{\theta}\delta
v^{\hat r}+ \frac{N}{\sin^2\theta}\partial_{r}\partial_{\phi}S_0
\,\partial_{\phi}\delta v^{\hat r}\Bigg] Y_{\ell'
m'}^{*}(\theta,\phi) d \Omega |_{r=R}. \
\end{eqnarray}

In the case of toroidal oscillations of a NS in the mode $(\ell',m')$, 
the oscillation velocity components are given by (\ref{v_toroidal}). 
Substituting the right-hand sides of (\ref{v_toroidal}) and 
(\ref{S_0}) into (\ref{deltaplm4}), we get the following expression 
for $\partial_t\delta s_{\ell m}(t)$ for a NS with a dipole magnetic 
field
\begin{eqnarray}
\label{dslm}
\partial_t\delta s_{\ell m}(t)&=&\frac{1}{\ell\left(\ell+1\right)}
\frac{B_{0}f_{_{R}}M^3 \tilde{\eta}_{_{\rm R}}}{R\,q_{\ell}
\left(R\right)}
\\\nonumber
&\times&\int_{4\pi}\left[\partial_{\theta} Y_{\ell
m}\left(\sin\theta\cos\chi-e^{i\phi}\cos\theta\sin\chi \right)
+ie^{i\phi}\partial_{\phi}Y_{\ell m}\sin\theta\sin\chi
\right]\frac{Y^*_{\ell' m'}(\theta,\phi)}{\sin\theta}d\Omega\ ,
\end{eqnarray}
 where $\tilde{\eta}_{_{\rm R}}=\eta_{_R} e^{-i\omega_{_{\rm R}}t}$. 
Substituting this into (\ref{exp_dS2}), we obtain the expression for 
$\delta S$ for a toroidal oscillation mode $(\ell',m')$ of a NS a with 
dipole magnetic field
\begin{eqnarray}
\label{dS}
\partial_t\delta S(r,t)&=&\sum_{\ell=0}^{\infty}\sum_{m=-\ell}^{\ell}
\frac{B_{0}R f_{_{R}}\tilde{\eta}_{_{\rm
R}}}{\ell\left(\ell+1\right)} \frac{r^2\,q_{\ell}
\left(r\right)}{R^2\,q_{\ell} \left(R\right)}
\\\nonumber
&\times&\int_{4\pi}\left[\partial_{\theta} Y_{\ell
m}\left(\sin\theta\cos\chi-e^{i\phi}\cos\theta\sin\chi \right)
+ie^{i\phi}\partial_{\phi}Y_{\ell m}\sin\theta\sin\chi
\right]\frac{Y^*_{\ell' m'}(\theta,\phi)}{\sin\theta}d\Omega\ .
\end{eqnarray}

If the magnetic dipole moment is aligned with the oscillation mode 
axis ($\chi=0$), one can easily show that
\begin{equation}
\label{deltaplm6}
\partial_t\delta s_{\ell m}(t)=\frac{im}{\ell\left(\ell+1\right)}
B_{0} R f_{_{R}} M^3\tilde{\eta}_{_{\rm R}}
\frac{\delta_{\ell'
\ell}\delta_{m'm}}{R^2\,q_{\ell}\left(R\right)}\ ,
\end{equation}
\begin{equation}
\label{dS0}
 \partial_t\delta S=\frac{\imath m'}{\ell'(\ell'+1)}
     B_{0} R f_{_{R}} \tilde{\eta}_{_{\rm R}}
\frac{r^{2}\,q_{\ell'}(r)}{R^2\,q_{\ell'}(R)}\,Y_{\ell' m'}\ ,
\end{equation}
 where $\delta_{\ell' \ell}$ is the Kronecker tensor.

\section{Solution of the $\Psi_{_{\rm SC}}$ Equation for $\chi=0$}
\label{section:solution_for_chi_eq_0}

Substituting the right hand side of equation (\ref{dS0}) and 
expressions (\ref{v_toroidal}) into (\ref{bound_Psi_dip}), we obtain a 
boundary condition for the function $\Psi_{_{\rm SC}}$ for toroidal 
oscillations of a NS, for the case $ \chi = 0 $:
\begin{equation}
\label{bound_tor} \Psi_{_{{\rm SC}}}|_{r=R}=- B_{0} R f_{_{
R}}\tilde{\eta}_{_{\rm R}}\int_{0}^{\theta} \left[\cos\vartheta\,
\partial_{\vartheta} Y_{\ell'
m'}(\vartheta,\phi)-\frac{m'^2}{\ell'\left(\ell'+1\right)}\frac{Y_{\ell'
m'}(\vartheta,\phi)}{\sin\vartheta}\right]d\vartheta\ ,
\end{equation}
 where $\vartheta$ is the integration variable. Substituting the 
expression for $\partial_t\delta S$ given by (\ref{dS0}) into equation 
(\ref{Psi_SC}), we obtain an equation for the SC potential, 
$\Psi_{_{\rm SC}}$, for a toroidal oscillation mode $(\ell',m')$ of a 
NS having a dipole magnetic field aligned with the oscillation mode 
axis
\begin{equation}
\label{Psi_SC_fin_m} -2r^2
q_1(r)\cos\theta\,\partial_r(\Psi_{_{\rm SC}})+\partial_r [r^2
q_1(r)]\sin\theta\,\partial_{\theta}(\Psi_{_{\rm
SC}})-\frac{m'^{2}}{\ell'(\ell'+1)} B_{0}R f_{_{
R}}\tilde{\eta}_{_{\rm R}} \partial_r [r^2 q_1(r)]\frac{r^2
q_{\ell'}(r)}{R^2 q_{\ell'}(R)}\, Y_{\ell' m'}(\theta,\phi )= 0\ .
\end{equation}

This is a first-order partial differential equation. According to a 
well-known theorem from the theory of such equations, 
(\ref{Psi_SC_fin_m}) is equivalent to the following system of 
first-order ordinary differential equations (see, for example, Chapter 
II of Volume II of \citet{courant_62} for a thorough discussion)
\begin{equation}
\label{system} -\frac{dr}{2r^2
q_1(r)\cos\theta}=\frac{d\theta}{\partial_r [r^2
q_1(r)]\sin\theta} = \frac{d\Psi_{_{\rm
SC}}}{\frac{m'^{2}}{\ell'(\ell'+1)}B_{0}R f_{_{
R}}\tilde{\eta}_{_{\rm R}} \partial_r [r^2 q_1(r)]\frac{r^2
q_{\ell'}(r)}{R^2 q_{\ell'}(R)}\, Y_{\ell' m'}(\theta,\phi
)}=\frac{d\phi}{0}=\frac{dt}{0}\ .
\end{equation}
 The characteristics of this system are
\begin{eqnarray}
\label{char1}
&& \varphi_{0}=t \ ,\\
&& \varphi_{1}=\phi \ ,\\
\label{char2} && \varphi_{2}=\sqrt{-r^2 q_1(r)}\sin\theta\ , \\
\label{char3} && \varphi_{3}=\Psi_{_{\rm SC}}+
\frac{1}{2}\frac{m'^{2}}{\ell(\ell+1)} B_{0}R f_{_{
R}}\tilde{\eta}_{_R}\,\int \frac{\partial_r [r^2 q_1(r)]}
{q_1(r)}\frac{q_{\ell'}(r)}{R^2 q_{\ell'}(R)} \,\frac{Y_{\ell' m'}
(\theta(r) ,\phi)}{\cos\theta (r)}\,dr\ ,
\end{eqnarray}
 where $\theta(r)$ depends on the variable $r$ due to (\ref{char2}):
\begin{equation}
\label{theta_r}
\theta\left(r\right)=\arcsin\left(\frac{\varphi_{2}} {\sqrt{-r^2
q_1(r)}}\right)\, .
\end{equation}
 The integral of equation (\ref{Psi_SC_fin_m}) is an arbitrary 
function of $\varphi_{0},\,\varphi_{1},\,\varphi_{2}$ and 
$\varphi_{3}$:
\begin{equation}
\label{integ}
\Gamma(\varphi_{0},\varphi_{1},\varphi_{2},\varphi_{3})=0\ .
\end{equation}
 Using this, one can obtain an expression for the general
solution of equation (\ref{Psi_SC_fin_m}):
\begin{equation}
\label{gen_sol}
    \Psi_{_{\rm SC}}=-\frac{1}{2}
    \frac{m'^{2}}{\ell'(\ell'+1)}
    B_{0}R f_{_{ R}}\tilde{\eta}_{_R}\,\int\frac{\partial_r [r^2 q_1(r)]}
    {q_1(r)}\frac{q_{\ell'}(r)}{R^2 q_{\ell'}(R)}
      \,\frac{Y_{\ell' m'}(\theta
    (r) ,\phi)}{\cos\theta (r)}\,dr
    +\Phi_{1}\left[\sqrt{-r^2 q_1(r)}\,\sin\theta,
      \phi,t\right]\ ,
\end{equation}
 where $\Phi_1$ is an arbitrary function that has to be determined 
from the boundary condition (\ref{bound_tor}). Using 
(\ref{bound_tor}), we can obtain a boundary condition for $\Phi_1$:
\begin{equation}
\label{bound_Phi} \Phi_{1}\left[\sqrt{-r^2 q_1(r)}\,\sin\theta,
  \phi,t\right]|_{r=R}=
\Psi_{_{\rm SC}}|_{r=R}+\frac{1}{2}
    \frac{m'^{2}}{\ell'(\ell'+1)}
    B_{0}R f_{_{ R}}\tilde{\eta}_{_R}\,\int\frac{\partial_r [r^2 q_1(r)]}
    {q_1(r)}\frac{q_{\ell'}(r)}{R^2 q_{\ell'}(R)}
      \,\frac{Y_{\ell' m'}(\theta
    (r) ,\phi)}{\cos\theta (r)}\,dr|_{r=R}\ ,
\end{equation}
 where $\Psi_{_{\rm SC}}|_{r=R}$ is given by (\ref{bound_tor}). Now 
using equations (\ref{gen_sol}) and (\ref{bound_Phi}), we can write a 
final expression for the general solution of equation 
(\ref{Psi_SC_fin_m}) in the following form
\begin{equation}
\label{gen_sol_1_app} \Psi_{_{\rm SC}}=-\frac{1}{2}
    \frac{m'^{2}}{\ell'(\ell'+1)}
B_{0}R f_{_{ R}}\tilde{\eta}_{_R}\, \int_{R}^{r}
\frac{\partial_{r'} [r'^2 q_1(r')]}{q_1(r')}\frac{q_{\ell'}(r')}
{R^2 q_{\ell'}(R)}\,\frac{Y_{\ell' m'}(\theta(r'),\phi)}
{\cos\theta (r')}\,dr' +\Phi_{2}\left[\sqrt{-r^2 q_1(r)}
\,\sin\theta,\phi,t\right]\ ,
\end{equation}
 where $r'$ is the integration variable. The function $ \theta(r) $ 
under this integral must be substituted by (\ref{theta_r}) and, after 
performing the integration, the function $ \phi_2 $ should be replaced 
by (\ref{char3}). The unknown function $ \Phi_{2} \left[ \sqrt{-r^2 
q_1(r)} \, \sin \theta, \phi, t \right] $ is determined using boundary 
condition (\ref{bound_tor}):
\begin{equation}
\label{phi2} \Phi_{2}\left[\sqrt{-r^2
q_1(r)}\,\sin\theta,\phi,t\right]|_{r=R}=-B_{0} R f_{_{
R}}\tilde{\eta}_{_R}\int_{0}^{\theta} \left[\cos\vartheta\,
\partial_{\vartheta} Y_{\ell'
m'}(\vartheta,\phi)-\frac{m'^2}{\ell'\left(\ell'+1\right)}\frac{Y_{\ell'
m'}(\vartheta,\phi)}{\sin\vartheta}\right]d\vartheta\ .
\end{equation}
 In order to obtain $ \Phi_{2} $ for arbitrary $r$, one has to express 
the right hand side of this equation in terms of $\sin\theta$, and 
then replace $\sin\theta$ by $ \sqrt{ [r^2 q_1(r)] / [R^2 q_1(R)] } 
\times \sin \theta $. We then have our analytical expression for the 
function $\Psi_{_{\rm SC}}$ for toroidal oscillations of a NS. 

\section{Energy Losses for $ m' \neq 0 $ Modes}
\label{section:energy_loss_m_ne_0}

In this Section we solve equation (\ref{Psi_SC_fin_m}) and calculate
$ \theta_0 $ together with the total energy losses for
non-axisymmetric toroidal oscillation modes for small $ \theta_0
$. Taking the limit of small $ \theta $ in the boundary condition
(\ref{bound_tor}) for $\Psi_{_{\rm SC}}$, we obtain the following
approximate expression for $ \Psi_{SC}|_{r=R} $
\begin{equation}
\Psi_{SC}|_{r=R}=-B_0 R f_{_R}
\tilde\eta_{_R}\left\{\left[1-\frac{m'}{\ell'(\ell'+1)}\right]
A^{(1)}_{\ell' m'} \theta^{m'} + \left[ A^{(2)}_{\ell' m'}
- \frac{m'A^{(1)}_{\ell' m'}}{2(m'+2)}-\frac{m'^2 \left(A^{(1)}_{\ell'
    m'}/6+A^{(2)}_{\ell' m'}\right)}{(m'+2)\ell'(\ell'+1)}
\right] \theta^{m'+2} \right\} .
\end{equation}
 Using this result, one can show that the general solution of equation 
(\ref{Psi_SC_fin_m}) for small $\theta$ has the following form
\begin{eqnarray}
\label{eqq:geneneral_solution}\nonumber
\Psi &=& -B_0 R f_{_R} \tilde\eta_{_R}g_1(r)^{m/2}\bigg\{
\left[\frac{m^2}{\ell'(\ell'+1)}  \int_R^r
\frac{g_1'(r)}{g_1(r)^{m/2+1}}
\frac{g_{\ell'}(r)}{g_{\ell'}(R)}dr+\frac{1}{g_1(R)^{m'/2}}
\left[1-\frac{m'}{\ell'(\ell'+1)}\right]
\right]A^{(1)}_{\ell' m'} \theta^{m'} \\\nonumber\\\nonumber
& + & \bigg[\left[\frac{m'+3}{6}A^{(1)}_{\ell' m'}+A^{(2)}_{\ell' m}
\right]g_1(r) \int_R^r \frac{g_1'(r)}{g_1(r)^{m'/2+2}}
\frac{g_{\ell'}(r)}{g_{\ell'}(R)}dr -
\frac{m' A^{(1)}_{\ell' m}}{6} \int_R^r \frac{g_1'(r)}
{g_1(r)^{m'/2+1}} \frac{g_{\ell'}(r)}{g_{\ell'}(R)}dr \\\nonumber\\
&+& \frac{1}{g_1(R)^{m'/2}}\left[A^{(2)}_{\ell' m'} -
\frac{m'}{2(m'+2)}A^{(2)}_{\ell' m'} -\frac{m'^2}{(m'+2)\ell'(\ell'+1)}
\left(\frac{1}{6}A^{(1)}_{\ell' m'} + A^{(2)}_{\ell' m'}\right)\right]
\frac{g_1(r)}{g_1(R)}\\\nonumber\\\nonumber
&+&\frac{m'}{6}\left[1-\frac{m'}{\ell'(\ell'+1)}\right]
\left(\frac{g_1(r)}{g_1(R)}-1\right)
\bigg] \theta^{m'+2} \bigg\} \ ,
\end{eqnarray}
 where we have introduced a new function $ g_{\ell}(r) = r^2 
q_{\ell}(r) $ for simplicity of notation. Using (\ref{ex_ef2}) and 
(\ref{eqq:particle_energy}), we obtain an expression for $\Delta {\cal 
E}$:
\begin{equation}
\label{eqq:delta_E_theta}
\Delta {\cal E} = - B_0 R f_{_R} \tilde\eta_{_R} N_{_R} \int_0^{\theta} \cos
  \theta \partial_{\theta} Y_{\ell' m'} d \theta \ .
\end{equation}
 Substituting the expansion of $ Y_{ \ell' m' } $ for small $\theta$
given by (\ref{expan_sh}) into (\ref{eqq:delta_E_theta}), we get the
following expression for $\Delta {\cal E}$ for small $\theta$
\begin{equation}
\label{eqq:delta_E_small_theta}
\Delta {\cal E} = - B_0 R f_{_R} \tilde\eta_{_R} N_{_R} \left[
    A^{(1)}_{\ell' m'} \theta^{m'} + \left( A^{(2)}_{\ell' m'} -
    \frac{m'}{2(m'+2)} A^{(1)}_{\ell' m'} \right) \theta^{m'+2}\right] \ .
\end{equation}
 Now substituting (\ref{eqq:geneneral_solution}) into 
(\ref{SC_charge}), we get an expression for $\rho_{_{\rm 
SC}}(R,\theta,\phi)$:
\begin{equation}
\label{eqq:rho_sc_small_theta}
\rho_{_{\rm SC}}(R,\theta,\phi)= - \frac{B_0 f_{_R} \tilde\eta_{_R}}
    {\pi N_{_R} R} D_{\ell' m'} (R,M) \theta^{m'} \ ,
\end{equation}
 where we have introduced a new quantity $D_{\ell'  m'}(R, M)$ for
simplicity of notation:
\begin{eqnarray} \nonumber
  \label{eqq:d_lm}
  \
  D_{\ell' m'}(R,M) & = & \frac{m'}{48} \bigg\{ 4 A^{(1)}_{\ell' m'} (1 +
  m') \left[ 1 - \frac{m'} { \ell' ( \ell' + l ) } \right] +
  \frac{ 3 } { \ell' (\ell' + 1) } \bigg\{ 8 m' (1 + m') \left[
    \frac{ A^{ (1) }_{ \ell' m' } m' }{ 2 (2 + m') } +
    \frac{ (A^{ (1) }_{ \ell' m'} + 6 A^{ (2) }_{ \ell'
        m' } ) m'^2 } { 6 (\ell' + \ell'^2 ) (2 + m') } - A^{ (2) }_{
      \ell' m' } \right] \\\nonumber\\ & + & \frac{
    A^{(1)}_{ \ell' m' } ( 2 M - R ) } { g_1(R)^2 g_{\ell'}(
    R) } \bigg[ \partial_r g_1(r) \ \big( \ell' (\ell' +
    l) \ g_{\ell'}(R) (4 g_1(R) + ( m' - 2 ) R
    \partial_r\ g_1(r) + 2 m R g_1(R) \partial_r g_{\ell'}(r)\big)
    \\\nonumber\\\nonumber & + & 2 \ell' ( \ell' +
    l ) R g_1(R) g_{\ell'}(R) \partial^2_{r, r} g_1( r )
    \bigg] \bigg\} \bigg\} |_{ r = R } \ . \nonumber
\end{eqnarray}
 Substituting (\ref{eqq:delta_E_small_theta}) and 
(\ref{eqq:rho_sc_small_theta}) into (\ref{eqq:theta_0}), we obtain an 
algebraic equation for $\theta_0$ for $ m \neq 0 $ toroidal 
oscillation modes, which has the solution
\begin{equation}
\label{eqq:theta_0_mne0}
\theta_0 = \left[ \frac{16 N_R \ \tilde\eta_{_R}^2 | A^{(1)}_{ \ell' m' }
D_{\ell' m'} (R,M) | }{ f_{_R}^2 } \right]^{\frac{1}{6-2m'}} \ .
\end{equation}
 We can see from this expression that $\theta_0$ is small for modes 
with $m' < 3$, in agreement with the estimate of TBS. Remarkably, 
$\theta_0$ does not depend on $B_0$ as in the case of $m'=0$. 
Substituting (\ref{eqq:delta_E_small_theta}), 
(\ref{eqq:rho_sc_small_theta}) and (\ref{eqq:net_cur}) into 
(\ref{eqq:en_loss}) gives
\begin{equation}
L_{\ell' m'}=\frac{2}{(m'+1)\pi} \left( B_0 R f_{_R} \tilde \eta_{_R}
    \right)^2 \ | A^{(1)}_{ \ell' m' } \ D_{\ell' m'} (R,M) | \
    \theta_0^{2m'+2} .
\end{equation}
Replacing $\theta_0$ in this equation by the right hand side of
equation (\ref{eqq:theta_0_mne0}), we obtain 
\begin{equation}
\label{eqq:energy_loss_mne0}
  L_{\ell' m'}=\frac{B_0^2 R^2 f_{_R}^4 N_R^{\frac{1+m'}{3-m'}}}{8\pi(m'+1)}
  \left[ \frac{16 \tilde\eta_{_R}^2 \ | A^{(1)}_{ \ell' m' } \
  D_{\ell' m'} (R,M) | }{ f_{_R}^2 } \right]^{\frac{4}{3-m'}} \
  \ .
\end{equation}

\label{lastpage}

\end{document}